\documentclass[%
 reprint,
showpacs,preprintnumbers,
 amsmath,amssymb,
 aps,
]{revtex4-1}

\usepackage{graphicx}
\usepackage{dcolumn}
\usepackage{bm}

\usepackage[pagebackref=false,colorlinks,linkcolor=blue,filecolor=green,urlcolor=blue ,citecolor=blue]{hyperref}
\usepackage{soul}

\usepackage{xcolor}

\begin{document}

\title{Electronic correlations in magnetized helical edge states coupled to s-wave superconductors}
\author{Zeinab Bakhshipour and Mir Vahid Hosseini}
 \email[Corresponding author: ]{mv.hosseini@znu.ac.ir}
\affiliation{Department of Physics, Faculty of Science, University of Zanjan, Zanjan 45371-38791, Iran}

\date{\today}

\begin{abstract}
We theoretically study the role of electron-electron interactions in one-dimensional magnetized helical states coupled to an s-wave superconductor. We consider a partially mixed helical (superhelical) regime, where the magnetic field (superconductivity) has a dominant contribution. Using bosonization and renormalization group techniques, it is shown how the interactions affect the correlation functions in the system. In the partially mixed helical regime, we find that superconducting pairing promotes spin-density-wave correlations, while singlet and triplet pairing correlations suppress, especially under attractive interactions. In contrast, in the superhelical regime, a perturbative Zeeman field enhances both spin singlet and spin triplet-x pairings under repulsive interactions. We calculate both logarithmic and residual corrections to charge-density-wave, spin-density-wave, and pairing correlations, revealing short- and long-range behaviors. We further investigate spin transport properties supplemented by a renormalization group analysis of the temperature and frequency dependence of spin conductivity. We also analyze the transport of momentum-spin-locked carriers in the presence of Zeeman-induced and pairing-induced gaps, uncovering the effect of interactions and the interplay between the two gaps in helical systems.
\end{abstract}

\maketitle

\section {Introduction} \label{s1}
Topological quantum computation has been one of the most exciting topics in condensed matter physics in recent decades due to its immuneness to local perturbations \cite{Kitaev2003,Nayak2008}. Majorana zero modes are promising candidates for realizing the required platforms  \cite{Kitaev2001,Alicea2012}, which are exotic quasiparticles that obey non-Abelian statistics and can be emerged in topological superconductors \cite{Beenakker2013,sato2016}. The nonlocal feature of Majorana modes makes them robust against local sources of perturbations, allowing operations that are topologically protected \cite{Microsoft2025}. The first quantum processing unit based on topological superconducting qubits has been demonstrated by Microsoft’s Station Q group \cite{Microsoft2025,Microsoft,Garisto}.

Meanwhile, conventional superconductors that have weak spin-orbit interactions, as well as intrinsic topological superconductors such as odd-parity pairing \cite{Ruhman2017,Maneo1994,Hor2010,Hsieh2012}, or superconductors without a center of symmetry that do not have a specific parity \cite{Sato2009,Tanaka2009,Bauer2004,Badica2005}, will not contribute to hosting Majorana modes. The idea of inducing topological superconductivity in hetero structures was formed by Fu and Kane \cite{Fu2008}. In this regard, an s-wave superconductor is placed near the topological surface states of a 3D topological insulator, during which the helical spin polarization \cite{Sato2009} of the surface states manifests a 2D effective p-wave superconductor \cite{Fu2008,Sato2010,Sau2010,Alicea2010} hosting Majorana states \cite{Lutchyn2010}. This issue has been extensively studied in many experimental works \cite{Williams2012,Wang2012,Yang2012,Wang2013,Cho2013,Oostinga2013,Finck2014,Snelder2014}. Moreover, recent experiments \cite{Mourik2012,Deng2016,Albrecht2016,Zhang2018,Nichele2017,Vaitiekenas2020,Aguado2017} have reported the evidence for Majorana modes in fabricated metal-superconductor or semiconductor-superconductor hybrid systems that host strong spin-orbit interactions \cite{Manchon2015}.

One-dimensional (1D) topological superconductivity can be engineered in systems hosting helical states that are proximitized by an s-wave superconductor in the presence \cite{Fu2008,Sato2009,Tanaka2009,Lutchyn2010} or absence \cite{Orth2015,Pedder2017} of a Zeeman field. The 1D helical states may arise as edge states in two-dimensional quantum spin Hall insulators \cite{Kane2005,Koing2007a,Wu2006,Berneving2006,Qi2010,SpinHall}. Furthermore, they can be found in quasi-1D systems with strong spin-orbit coupling, including semiconductor nanowires \cite{Lutchyn2010}, carbon nanotubes \cite{ Klinovaja2011, Klinovaja2012a,Izumida2017,Marganska2018}, bilayer graphene \cite{Klinovaja2012b}, and graphene nanoribbons \cite{Klinovaja2013a}. In these systems, the interplay of spin-orbit coupling, magnetic field, and proximity-induced superconductivity can establish topological superconductivity. Various setups have been proposed to probe topological superconductivity, including Josephson junctions \cite{Deacon2017,Pedder2017,Udupa2021}, superconductor–helical boundary lines \cite{Hart2014, Klinovaja2013b}, and hybrid structures connecting multiple edge states~\cite{Qingyun2017,Fleckenstein2021, Eriksson2015}. 

In one dimension, electron-electron interactions give rise to the Luttinger liquid behavior \cite{1Dreview,Imambekov2012}, where correlation functions exhibit power-law scaling with exponents determined by the interaction strength \cite{Giamarchi2004}. The study of quantum correlation functions has been extensively investigated in many research areas \cite{LLRange} ranging from spin chains \cite{Mahdavifar2015} to topological systems in the presence of superconductivity \cite{Ruhman2017}. The role of strong interactions in quasi-helical Majorana edge states \cite{Gangadharaiah2011} and in controlling the destructive effect of magnetic fields on superconductors has been pursued \cite{Klinovaja2014,Thakurathi2018}. Also, more exotic quasi-particles have been discovered \cite{Fendley2012,Orth2015}, in systems with strong electron interactions and a pairing mode such as Cooper-pair tunneling into a quantum Hall system \cite{Michelsen2020}, due to the proximity effect of a superconductor with helical quantum wires \cite{Michelsen2023}. The analysis of correlations is a powerful tool for understanding the behavior of dominant excitations in these systems, paving the way for future advances in the fields of spintronics \cite{Data1990,Fabian2007}, spin-orbitronics \cite{Fabian2007,Chappert2007}, and quantum computing \cite{Hutter2016}.

The relatively long coherence length and considerable relaxation time in superconducting systems make them ideal platforms for spin-based quantum manipulation. Additionally, the study of quantum spin phases \cite{Mahdavifar2024} and of helical edges subject to time-reversal-symmetry-breaking perturbations has opened new routes toward controlling spin–spin correlations in partially mixed helical (PMH) states \cite{Hosseini2020,Bakhshipour2024}. In these states, a transverse magnetic field or local magnetic moments rotate the spin quantization axis of the helical edge modes, so that spin and momentum are no longer perfectly locked, as illustrated for helical edges in a magnetic field \cite{Soori2012} and for helical edges gapped by magnetic impurities \cite{Wozny2018}. The resulting PMH regime, where spin–momentum locking is only partial, provides an interesting playground to explore the interplay of superconductivity, magnetism, and interactions. Motivated by these perspectives, it is interesting to investigate the behavior of spin and charge correlations in a 1D topological superconductivity formed at the interface between an s-wave superconductor and a helical edge state subject to a Zeeman field \cite{Qi2011}.

In this work, we employ bosonization and renormalization group (RG) techniques to study the interaction effects in a 1D magnetized helical state coupled to an s-wave superconductor. We consider two complementary regimes: (i) a PMH state with a dominant Zeeman gap perturbed by proximity-induced superconductivity and (ii) a superhelical state with dominant superconducting pairing perturbed by a transverse Zeeman field. In both cases, we analyze many body quantum correlation functions such as the charge-density-wave (CDW), spin-density-wave (SDW), and superconducting pairing correlations. We also identify the leading instabilities in different interaction regimes for short- and long-range contributions.

Our results reveal that, under repulsive interactions, the perturbative Zeeman field can increase the superconducting correlations, particularly the triplet-x component. However, singlet pairing has negligible values. In contrast, attractive interactions stabilize singlet superconductivity while also amplifying triplet-x correlation as the leading phase. The competition between magnetic and superconducting gaps leads to non-trivial corrections to correlation functions, which we calculate both perturbatively and non-perturbatively. Furthermore, combining both the Kubo formalism and the Memory function method, we evaluate the frequency and temperature dependence of the spin conductivity and show how transport properties reflect the underlying correlated phases.

The paper is organized as follows. In Sec. \ref{II}, we introduce the model system. Using a bosonization approach, gapped forms corresponding to both the Zeeman and superconducting proximity terms are derived. In Sec. \ref{III}, we perform a RG analysis to examine the interplay between electron-electron interactions and the gap parameters. Section \ref{IV} is devoted to the analysis of charge-spin density wave (CDW/SDW) and superconducting pairing correlation functions in various regimes of interaction and gap strengths. In Sec. \ref{V}, we investigate corrections to these correlation functions in the strong, intermediate, and weak gap regimes, including logarithmic or perturbative approaches. In Sec. \ref{s6},  the spin transport properties of the system is investigated, focusing on the frequency and temperature dependence of the spin conductivity using the Memory function approach and RG-improved results. Finally, we summarize our main findings in Sec. \ref{s7}.
\begin{figure}[t!]
    \centering
    \includegraphics[width=8cm]{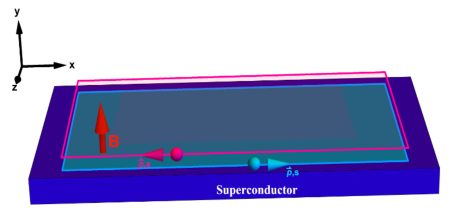}
   \caption{(Color online) Schematic of a helical edge of a quantum spin Hall insulator exposed to a Zeeman field and adjacent to an s-wave superconductor forming a topological superconductor. }
    \label{fig1}
\end{figure}

\section {Model}\label{II}

We consider a continuum model of a quantum helical state, described by Hamiltonian $\mathcal{H}_{hel}$ \cite{Koing2007a,Wu2006}, subject to a transverse Zeeman field, with Hamiltonian $\mathcal{H}_z$ \cite{Hosseini2020,Soori2012,Wozny2018}. Additionally, electrons experience a proximity-induced pairing potential generated by a conventional s-wave superconductor, described by $\mathcal{H}_s$ \cite{Alicea2012}. So, the system, shown in Fig. (\ref{fig1}), has the total Hamiltonian, $\mathcal{H}=\mathcal{H}_{hel}+\mathcal{H}_{z}+\mathcal{H}_{s}$, where
\begin{align}
\mathcal{H}_{hel} &= v_F \, \psi^{\prime\dagger}(x) \, k_x \sigma_x \, \psi^\prime(x),
\label{Ham1}\\
\mathcal{H}_z &= \Delta_z \, \psi^{\prime\dagger}(x) \sigma_y \psi^\prime(x),\label{Ham1Zeem} \\
\mathcal{H}_s &= \Delta_s \, \psi_L^\prime(x) \psi_R^\prime(x) + \mathrm{h.c.}. \label{Ham1Sup}
\end{align}
Here, $v_F$ is the Fermi velocity, $k_x$ is the wave vector along the wire, and $\sigma_{x,y}$ are the Pauli matrices acting in spin space. $\Delta_z$ and $\Delta_s$ denote the Zeeman and superconducting pairing gaps, respectively.

The single-electron field operator $\psi^\prime(x)$ can be expanded in terms of right- ($r=R$) and left-moving ($r=L$) components as
\begin{equation}
\psi^\prime(x)=\sum_{r=R,L}\psi^\prime_r(x),
\label{fieldoperator}
\end{equation}
where the chiral components are $\psi^\prime_r(x)=\chi_r(r k_F) \, \psi_r(x) e^{i r k_F x}$. The helical eigenmode $\psi_r(x)$ annihilates an electron moving in direction $r$ at position $x$ with Fermi momentum $k_F$. The spinor $\chi_r(k)$, containing the spin structure, is the eigenstate of the magnetized helical state, which is given by
\begin{equation}
\chi_r(k) = \frac{1}{\sqrt{2}} \begin{pmatrix} r e^{i \vartheta_k} \\ e^{-i \vartheta_k} \end{pmatrix},
\end{equation}
where the mixing angle $\vartheta_k$ encodes the Zeeman-induced spin rotation,
\begin{equation}
\vartheta_k = \frac{1}{2} \arctan \left( - \frac{\Delta_z}{v_F k} \right).
\end{equation}
Note that in helical edge states spin and momentum are locked, so right- and left-moving modes have opposite spin projections. In this standard notation the spin index is often suppressed and the direction index R/L implicitly carries the spin information, so that $\psi_R \equiv \psi_{R\uparrow}$ and $\psi_L \equiv \psi_{L\downarrow}$.

To obtain the quasiparticle spectrum in the presence of both the Zeeman field and s-wave pairing, we construct the corresponding Bogoliubov–de Gennes Hamiltonian in Nambu space and diagonalize it.
For $\mu=0$ the eigenvalues can be written as
\begin{equation}
  \epsilon_{\pm,\eta}= \eta\, \sqrt{v_F^2 k_x^2 + (\Delta_z\pm\Delta_s)^2}, 
  \qquad \eta = \pm 1,
\end{equation}
which yields two effective single-particle gaps
$2 \bigl||\Delta_z|\pm|\Delta_s|\bigr|$ at the Fermi level. The smaller gap closes when $\Delta_z=\Delta_s$.

\begin{figure}[t!]
    \centering
    \includegraphics[width=9cm]{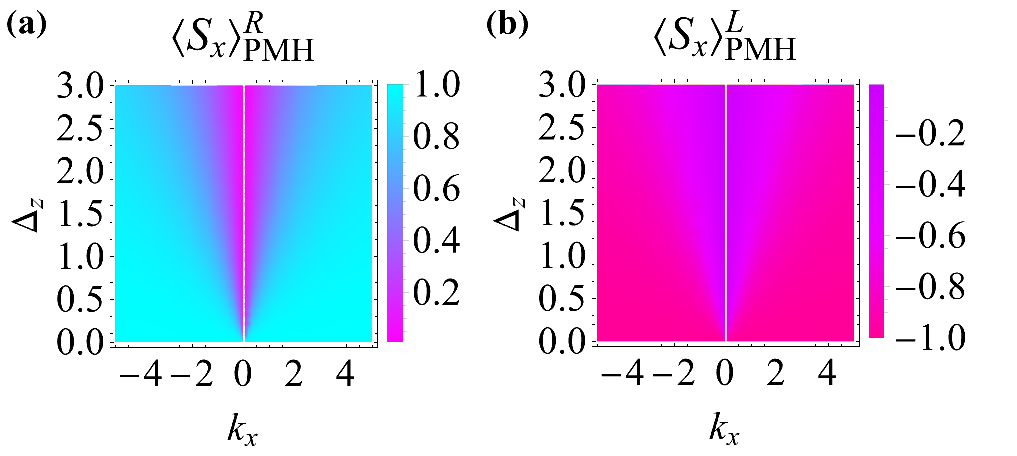}
   \caption{(Color online) Expectation values of $S_x$ for (a) the right  and (b) the left movers calculated with PMH states versus $\Delta_z$ and $k_x$. Increasing the Zeeman field raises the mixation of spin-x up and down around $k_x=0$. Here, $\Delta_s=0$.}
    \label{fig2-expectation}
\end{figure}

The Zeeman gap partially disrupts the pure longitudinal spin channel of the helical states, causing slight spin mixing of the right- and left-moving quasi-particles. Figure \ref{fig2-expectation} shows the expectation 
values of the x-component of helical electron spin, $S_x$, as functions of $\Delta_z$ and $k_x$ for the right movers (panel a) and the left movers (panel b). When the Zeeman gap $\Delta_z$ is zero, the spin expectation values of the right- and left-moving carriers are at the highest value in magnitude. This indicates that the spin quantization axis is parallel to the 1D conducting helical channel, i.e., the x-direction. For nonzero values of the Zeeman field, being in the y direction, the angle $\vartheta_k$ causes mixing between the different spin states near the helical edge, so that at $k_x=0$, the spin quantization axis of the helical wire rotates towards the direction of the Zeeman field. As $\Delta_z$ increases, the loss of spin alignment along the x-direction extends to larger momentum states resulting in the PMH state. We consider the Fermi level tuned close to the Dirac point, where both Zeeman and superconducting terms open low-energy gaps and govern the dynamics of the helical carriers.

We incorporate the effects of electron-electron interactions through the interaction Hamiltonian \cite{Voit1995},
\begin{align}
\mathcal{H}_{int} = &\, g_1 \, \psi_L^{\prime\dagger}(x) \psi_R^{\prime\dagger}(x) \psi_L^\prime(x) \psi_R^\prime(x) \nonumber \\
+ &\, g_2 \, \psi_L^{\prime\dagger}(x) \psi_L^\prime(x) \psi_R^{\prime\dagger}(x) \psi_R^\prime(x) \nonumber \\
+ &\, \frac{g_4}{2} \sum_{r=R,L} [\psi_r^{\prime\dagger}(x) \psi_r^\prime(x)]^2,
\end{align}
where $g_1$, $g_2$, and $g_4$ denote the backscattering, dispersive, and forward scattering interaction strengths, respectively. For a perfectly half-filled helical edge ($\mu=0$, $k_F=0$) one can in principle have an umklapp term, which becomes relevant only for strong repulsive interactions. In the present work we focus on weak-to-moderate interactions, and on a Fermi level tuned close to, but not exactly at, the Dirac point. Consequently, this umklapp term is RG-irrelevant or oscillatory and will be neglected in what follows.

Using the standard bosonization method, we express fermionic operators in terms of bosonic fields. So, the right- and left-moving fermionic operators can be represented as \cite{Tsvelik}
\begin{equation}
\psi_r^{(\dagger)}(x) = \frac{1}{\sqrt{2\pi a_0}} \, \eta_r \, e^{\pm i r \sqrt{4\pi} \phi_r(x)},
\end{equation}
where $a_0$ is a short-distance cutoff, $\phi_r$ are the chiral bosonic fields, and $\eta_r$ are Klein factors ensuring proper anticommutation relations. In this formalism, the full Hamiltonian $\mathcal{H} = \mathcal{H}_{hel} + \mathcal{H}_z + \mathcal{H}_s$ can be written as
\begin{equation}
\begin{split}
\mathcal{H}=\frac{v}{2}[\frac{1}{K}(\partial_x\Phi)^2+K(\partial_x\Theta)^2]&+\frac{\tilde{\Delta}_z}{2\pi a_0}e^{i\sqrt{4\pi}\Phi}+h.c.\\&+\frac{\Delta_s}{\pi a_0}\sin(\sqrt{4\pi}\Theta),
\label{Hamfirst0}
\end{split}
\end{equation}
where $\tilde{\Delta}_z$ is \cite{Bakhshipour2024}
\begin{equation}
\tilde{\Delta}_z = \Delta_z e^{-i \alpha x}, \quad \text{with} \quad \alpha = |k_F| K^2 - 2 k_F.
\label{KAlpha}
\end{equation}
Also, $\Phi = \phi_R + \phi_L$ and $\Theta = \phi_R - \phi_L$ are canonically conjugate bosonic fields satisfying
$[\Phi(x),\partial_{x'}\Theta(x')]=i\pi\delta(x-x')$. The Luttinger liquid parameter $K$ and the velocity $v$ are given by
\begin{equation}
K = \sqrt{\frac{1 - \frac{y_{bf}}{2} + \frac{y_4}{2}}{1 + \frac{y_{bf}}{2} + \frac{y_4}{2}}},
\label{Luttingerparameter}
\end{equation}
\begin{equation}
v = v_F \sqrt{\left(1 + \frac{y_4}{2}\right)^2 - \left(\frac{y_{bf}}{2}\right)^2},
\label{Luttingervelocity}
\end{equation}
satisfying $v K = v_F$. Here, we have defined $g_{bf} = g_2 - g_1$ and the dimensionless couplings $y_{bf/4}= \frac{g_{bf/4}}{4 \pi v_F}$. As already mentioned above, we work with a single helical channel. As a consequence the interacting edge is described by a single Luttinger parameter $K$, characterizing the combined charge–spin fluctuations of the helical liquid.
In contrast to a generic spinful two-channel Luttinger liquid, separate parameters $K_{\mathrm{charge}}$ and $K_{\mathrm{spin}}$ are not required in this effectively spinless helical model.

The Zeeman gap $\Delta_z$ and the superconducting gap $\Delta_s$ introduce sine-Gordon terms in the bosonized Hamiltonian (Eq.~\ref{Hamfirst0}). In a helical edge, these terms do not gap independent charge and spin sectors, because such sectors do not exist separately. Instead they act on the single pair of helical modes whose spin and momentum are locked. The Zeeman term tends to pin the bosonic field $\Phi$, while the pairing term pins its conjugate field $\Theta$, leading to competing tendencies that cannot be assigned to separate charge or spin channels. Because helicity ties spin rotations to particle–hole motion, both gaps affect the collective excitations of the helical liquid in an intrinsically mixed way. So their competition is governed by the Zeeman field, superconducting pairing, and electron–electron interactions. For clarity, we first analyze the case $\vartheta_k = 0$, which isolates the dominant structure of the gap-opening terms; the effects of partial spin mixing ($\vartheta_k \neq 0$) will be incorporated later. It should be noted that proximity-induced superconductivity modifies not only the pairing properties but also the spin texture of the helical carriers, which, in turn, influences the effective dynamics of the helical mode. As a result, the system may undergo a phase transition controlled by electron interactions, the strength of the Zeeman field, and superconducting pairing, as will be discussed in the following section.

\section{Renormalization Group analysis}\label{III}

In the presence of electron-electron interactions, the system parameters must be renormalized. This renormalization changes the relevance of the gap-opening terms from the Zeeman and superconducting couplings, thereby modifying the phase transition boundaries. The RG flow equations for the dimensionless helical Luttinger parameter $K(l)$ and the dimensionless coupling constants $\mathcal{Y}_z(l)$ and $\mathcal{Y}_s(l)$ are given by
\begin{align}
\frac{dK(l)}{dl} &= -\frac{1}{4} \mathcal{Y}_z^2(l) K^2(l) + \frac{1}{4} \mathcal{Y}_s^2(l),\label{KRG} \\
\frac{d\mathcal{Y}_z(l)}{dl} &= (2 - K(l)) \mathcal{Y}_z(l),\label{YzRG} \\
\frac{d\mathcal{Y}_s(l)}{dl} &= \left(2 - \frac{1}{K(l)}\right) \mathcal{Y}_s(l),\label{YsRG}
\end{align}
where $\mathcal{Y}_z$ and $\mathcal{Y}_s$ are the dimensionless Zeeman and superconducting gap parameters, respectively, defined as $\mathcal{Y}_{z,s} = \frac{\sqrt{2}\Delta_{z,s} a_0}{v}$.

The RG equations (\ref{KRG})-(\ref{YsRG}) describe how the Luttinger parameter $K$ and the gap amplitudes $\mathcal{Y}_z$ (Zeeman) and $\mathcal{Y}_s$ (superconductivity) evolve with the change of length scale $l$. From Eq.~(\ref{KRG}), one observes that the flow of $K$ is suppressed by the Zeeman gap, which enhances repulsive interactions (thus lowering $K$), whereas superconducting pairing tends to increase $K$, corresponding to effectively stronger attractive interactions. According to Eq. (\ref{YzRG}), the Zeeman gap $\mathcal{Y}_z$ increases under RG flow when $K < 2$, indicating a relevant perturbation that opens a magnetic gap in the spectrum. In contrast, $\mathcal{Y}_z$ becomes irrelevant under RG flow if $K > 2$. From Eq. (\ref{YsRG}), one can see that the superconducting gap $\mathcal{Y}_s$ is RG relevant when $K > \frac{1}{2}$, promoting superconductivity in regimes of strongly attractive interactions. Thus, the competition between these two gaps and their mutual feedback with the interaction parameter $K$ allow us to map the phase diagram of the system. Furthermore, we note that, unlike many previous studies \cite{Braunecker2009a,Braunecker2009b,Visuri2020,Citro2020,Giamarchi1991}, in our model both gap amplitudes remain finite even in the absence of interactions. This means that the gaps cannot act directly on each other. Instead, the presence of each gap alters $K$, and $K$ also controls the flow of both gaps.

\begin{figure}[t]
    \centering
    \includegraphics[width=0.45\textwidth]{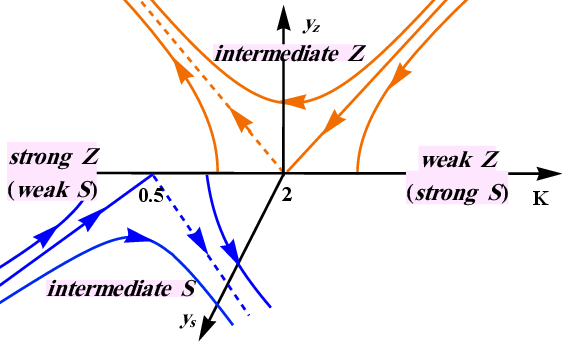}
    \caption{(Color online) Renormalization group flow trajectories for the dimensionless Zeeman gap $\mathcal{Y}_z$ (orange curves) and superconducting gap $\mathcal{Y}_s$ (blue curves) as a function of the Luttinger parameter $K$. The Zeeman coupling becomes relevant and flows to strong coupling for $K < 2$, while the superconducting pairing grows when $K > \frac{1}{2}$. The flow diagram illustrates distinct regimes characterized by dominant magnetic or superconducting orders, with an intermediate region where both gaps compete.}
    \label{fig:RGflows}
\end{figure}

Figure \ref{fig:RGflows} illustrates the RG flow trajectories, determined by numerical evaluation of Eqs. (\ref{KRG})-(\ref{YsRG}),  for the Zeeman and superconducting gaps as a function of the interaction parameter $K$. The Zeeman gap (orange curves) becomes relevant and flows to the strong gap regime at $K < 2$, signaling a magnetically dominated phase. In contrast, the superconducting gap (blue curves) increases when $K > \frac{1}{2}$, corresponding to a superconducting phase stabilized by attractive interactions. As a result, below the separatrices, where the Zeeman (superconducting) gap is strong and relevant, the superconducting (Zeeman) gap is weak and irrelevant. Moreover, there are intermediate regimes above the separatrices of magnetism and superconductivity where their gaps can compete and coexist, particularly for $1/2 < K < 2$. Thus, when one of the gaps (Zeeman or superconducting) is dominant or even moderate, the other can be treated as a perturbation. Consequently, the mutual effects of interactions with both magnetism and superconductivity in helical states can give rise to the rich phase diagram. 

Our RG equations describe the competition between Zeeman and superconducting gaps in an interacting helical edge tuned close to the Dirac point ($\mu \approx 0$), where the Zeeman operator is non-oscillatory and umklapp processes can be neglected. In this regime, the relevance conditions for the magnetic and pairing terms are directly analogous to those in the $\mu = 0$ analysis of Sela et al. \cite{Sela2011}  (and discussed more extensively in reference \cite{Alicea2012} and in other aspects in \cite{Fleckenstein2021,Klinovaja2013b,Lutchyn2010}), who obtained the global phase diagram of a strongly interacting helical liquid. In their work, the Zeeman and superconducting gaps appear as competing relevant perturbations with flow equations. The dependence on the Fermi level enters through the Fermi momentum $k_F$: at $k_F \approx 0$ our flows apply up to the largest length scales, while for $k_F\neq 0$ the Zeeman term acquires oscillatory factors and is effectively suppressed beyond a scale $\sim 1/|k_F|$. This incommensurate situation is analogous to the $\mu \neq 0$ case of Ref. \cite{Sela2011}, where the gapless line at zero magnetic and superconductivity with $\mu=0$ broadens into a gapless surface and a finite Zeeman field is required to open a gap.

Related analyses of interacting helical liquids and Majorana modes, using RG, exact scaling and numerical approaches, can be found for instance in Refs. \cite{Sarkar,Dey}, which also study the competition between Zeeman and superconducting gaps in interacting helical liquids and their mapping to spin-chain models. These works consider homogeneous helical liquids. Our results complement them by including the partially mixed helical edge regime and by connecting the RG flow more directly to the correlation functions and spin transport discussed in the following sections.

\section{Density and pairing correlation functions}
\label{IV}
In this section, we calculate the CDW and SDW correlation functions as well as the singlet (SS) and triplet (TS) superconducting pairing correlations for the helical states. In the following we use CDW, SDW, and singlet/triplet pairing correlation functions to identify the dominant ordering tendencies of the interacting edge in the different regimes introduced in Sec. \ref{III}. In a Luttinger liquid these correlation exponents are the natural quantities that encode which type of quasi–order (density, pairing, or spin) and which experimental probes (tunneling, Josephson, spin transport) are most strongly enhanced. Furthermore, we investigate how these correlations change in the presence of the combined pairing gap and Zeeman gap which enter the Hamiltonian through sine-Gordon terms arising from the superconductivity and Zeeman field to the helical state.

To proceed, we use the operators corresponding to the above-mentioned physical quantities \cite{Giamarchi1986,Giamarchi1988,Giamarchi1989},
\begin{align}
\mathcal{O}_{CDW}(x) &= \psi^{\prime \dagger}(x) \psi^{\prime}(x),
\label{CDW-op-equ}
\\
\mathcal{O}^{i}_{SDW}(x) &= \psi^{\prime \dagger}(x) \sigma_i \psi^{\prime}(x), \quad i = x, y, z,
\\
\mathcal{O}_{SS}(x) &= \psi^{\prime \dagger}_R(x) \psi^{\prime \dagger}_L(x),
\\
\mathcal{O}^{i}_{TS}(x) &= \psi^{\prime \dagger}_R(x) \sigma_i \psi^{\prime \dagger}_L(x), \quad i = x, y, z,
\label{TS-op-equ}
\end{align}
where $\psi^{\prime}(x)$ is the single-electron field operator defined in Eq. (\ref{fieldoperator}) and $\sigma_i$ are Pauli matrices acting in the spin space. We note that the spin modes of the right-moving and left-moving electrons correspond to spin-up and spin-down states, respectively. In the bosonization framework, the CDW, SDW, and superconducting pairing operators can be expressed in terms of the bosonic fields as,
\begin{align}
    \mathcal{O}_{CDW}(x,\tau)&=-\frac{1}{\sqrt{4\pi}}\partial_x\Phi(x,\tau),
 \label{CDWOp}
 \\\mathcal{O}^{x}_{SDW}(x,\tau)&=-\frac{1}{\sqrt{4\pi}}\partial_x\Theta(x,\tau),
  \label{xSDWOp}
  \\
 \mathcal{O}^{y}_{SDW}(x,\tau)&=\frac{1}{\pi a_{0}}\cos(\sqrt{4\pi}\Phi(x,\tau)+2k_{F}x),
   \label{ySDWOp}\\
 \mathcal{O}^{z}_{SDW}(x,\tau)&=-\frac{1}{\pi a_{0}}\sin(\sqrt{4\pi}\Phi(x,\tau)+2k_{F}x),
 \label{zSDWOp}\\\mathcal{O}_{SS}(x,\tau)&=\mathcal{O}^{x}_{TS}(x,\tau)=\frac{i}{2\pi a_0}e^{-i\sqrt{4\pi}\Theta(x,\tau)},\label{ss-pair-op}\\\mathcal{O}^{y}_{TS}(x,\tau)&=\mathcal{O}^{z}_{TS}(x,\tau)=0.\label{y,z-pair-op}
\end{align}

We use the definition of the density and pairing correlations as \cite{Giamarchi2004}:
\begin{align}
\mathcal{R}_{CDW}(r_1, r_2) &= \langle T_\tau \mathcal{O}_{CDW}(r_1) \mathcal{O}_{CDW}(r_2) \rangle, \label{RCDW} \\
\mathcal{R}^{ij}_{SDW}(r_1, r_2) &= \langle T_\tau \mathcal{O}^i_{SDW}(r_1) \mathcal{O}^j_{SDW}(r_2) \rangle, \\
\mathcal{R}_{SS}(r_1, r_2) &= \langle T_\tau \mathcal{O}_{SS}(r_1) \mathcal{O}_{SS}(r_2) \rangle, \\
\mathcal{R}^{ij}_{TS}(r_1, r_2) &= \langle T_\tau \mathcal{O}^i_{TS}(r_1) \mathcal{O}^j_{TS}(r_2) \rangle, \label{RTS}
\end{align}
where $T_\tau$ is the imaginary-time ordering operator, $r_1 = (x_1, v \tau_1)$, $r_2 = (x_2, v \tau_2)$, and $i,j = x,y,z$. In the following, we set $r_1 = r$ and $r_2 = 0$. Using the bosonized form of operators (\ref{CDWOp})–(\ref{y,z-pair-op}), the correlation functions with respect to $\mathcal{H}_{hel}$ can be expressed as
\begin{align}
\mathcal{R}_{CDW}(r) &= \frac{K}{4\pi^2} F_2(r), \label{CDWhel} \\
\mathcal{R}^{xx}_{SDW}(r) &= \frac{1}{4\pi^2 K} F_2(r), \label{xSDWhel} \\
\mathcal{R}^{yy}_{SDW}(r) &= \frac{\cos(\alpha x)}{2(\pi a_0)^2} e^{-2 K F_3(r)}, \label{ySDWhel} \\
\mathcal{R}^{zz}_{SDW}(r) &= \frac{1}{2(\pi a_0)^2} e^{-2 K F_3(r)}, \label{zSDWhel} \\
\mathcal{R}^{yx}_{SDW}(r) &= -\mathcal{R}^{xy}_{SDW}(r) = -\frac{\cos(\alpha x)}{2(\pi a_0)^2} e^{-2 K F_3(r)}, \label{yzSDWnonperturb} \\
\mathcal{R}_{SS}(r) &= \frac{1}{(2\pi a_0)^2} e^{-2 K^{-1} F_3(r)}, \label{ss-hel} \\
\mathcal{R}^{x}_{TS}(r) &= \frac{1}{(2\pi a_0)^2} e^{-2 K^{-1} F_3(r)}, \label{x-TS-hel} \\
\mathcal{R}^{y}_{TS}(r) &= 0, \label{y-TS-hel} \\
\mathcal{R}^{z}_{TS}(r) &= 0, \label{z-TS-hel}
\end{align}
where 
\begin{align}
F_2(r) &= \frac{(v \tau \, \mathrm{sign}(\tau) + a_0)^2 - x^2}{2 \left[ (v \tau \, \mathrm{sign}(\tau) + a_0)^2 + x^2 \right]^2}, \label{F2} \\
F_3(r) &= \frac{1}{2} \log \left( \frac{(v \tau \, \mathrm{sign}(\tau) + a_0)^2 + x^2}{a_0^2} \right). \label{F3}
\end{align}

In the following section, we incorporate the effects of superconducting pairing and Zeeman perturbations and calculate their impact on the correlation functions. In the weak-gap regime, these corrections are treated with two complementary approaches based on whether the operators contain exponential or non-exponential terms. For the exponential (long-range) components, we apply logarithmic corrections derived from the RG equations \cite{Giamarchi1989}. For the non-exponential terms (short-range), we use a perturbative expansion method up to second order in the gap parameters \cite{Citro2020}.

\section{Corrections to correlation functions in the strong, intermediate, and weak regimes}
\label{V}

As already discussed above, a weak gap in one channel corresponds to a strong gap in the other. We subsequently can evaluate corrections introduced by a weak pairing (Zeeman) gap to a helical wire with an intermediate Zeeman (pairing) gap in the following Sections (\ref{V-A} and (\ref{V-B})). This approach allows us to identify marginal boundaries for each gap in the weak regime while accounting for the presence of the relevant gap.

\subsection{Zeeman-induced PMH egde perturbed by superconductivity}
\label{V-A}

The presence of a superconductor near a helical state experiencing a Zeeman field can significantly modify the dominant fluctuations in the system. Here, we consider two cases: Strong Zeeman and intermediate Zeeman, and examine corrections introduced by superconducting perturbations.

\subsubsection{Strong Zeeman with Weak Pairing}
\label{V-A-1}

In the strong Zeeman regime, the bosonic field $\Phi$ becomes ordered, causing its dual field $\Theta$ to fluctuate rapidly. As a consequence, long-range correlations involving $\Theta$ vanish, suppressing pairing correlations. Following the calculations in Ref. \cite{Bakhshipour2024}, only the $x$- and $y$-components of the SDW survive, modified by the strong Zeeman gap as

\begin{align}
\mathcal{R}^{str-z,xx}_{SDW}(r) &= \frac{1}{4 \pi^2 K^*} F_2^*(r), \\
\mathcal{R}^{str-z,yy}_{SDW}(r) &= \frac{\cos(\alpha a_0)}{2 (\pi a_0)^2},
\end{align}

where

\[
F_2^*(r) = \frac{(v^* \tau \, \mathrm{sign}(\tau) + a_0)^2 - x^2}{2 \left[(v^* \tau \, \mathrm{sign}(\tau) + a_0)^2 + x^2\right]^2}.
\]

The $x$-component represents short-range correlations similar to the helical case but with renormalized Luttinger parameters $K^*$ and velocity $v^*$ affected by the Zeeman gap. The averaging over the pinned field $\Phi$ yields a constant value for the $y$-component, indicating its stability.

\subsubsection{Intermediate Zeeman with Weak Pairing: Zeeman-Affected Hamiltonian}
\label{V-A-2}

In this subsection we focus on the intermediate Zeeman, weak pairing regime, where the Fermi level lies in the magnetized band ($|\varepsilon_F| > \Delta_z$). So the low-energy edge is a gapless PMH Luttinger liquid with renormalized parameters $\tilde v$ and $\tilde K$. The Zeeman field is thus already encoded in these effective parameters and does not appear as an additional mass term at the Fermi points. On this Zeeman-affected background we then add a weak proximity pairing $\Delta_s$, which generates a sine-Gordon term for the dual field. The corresponding RG flow describes when this pairing becomes relevant within the PMH phase. In contrast to Sec. \ref{III}, we keep the Zeeman field fixed and only use the flow to determine how it shifts the interaction threshold for superconductivity.

We now consider the PMH state formed by an intermediate Zeeman field and study the effect of weak superconductivity. In
this case, the sine-Gordon-type Zeeman term can be approximated by a quadratic form, assuming the Zeeman
gap lies in a moderately relevant regime. Including the spin mixing angle $\vartheta_{k_F}$, the effective Hamiltonian reads
\begin{align}
\mathcal{H}_{PMH} &= \frac{\tilde{v}}{2} \left[ \frac{1}{\tilde{K}} (\partial_x \Phi)^2+ \tilde{K} (\partial_x \Theta)^2 \right]\nonumber\\ & + \frac{\Delta_s}{\pi a_0} \sin(2 \vartheta_{k_F}) \sin\left( \sqrt{4 \pi} \Theta \right),
\label{Hamfirst}
\end{align}
where the magnetized Luttinger parameter $\tilde{K}$ and velocity $\tilde{v}$ are given by
\begin{align}
\tilde{K} &= \sqrt{ \frac{ \tilde{v}_F - \frac{g_{bf}}{8\pi} + \frac{g_4}{8\pi} }{ \tilde{v}_F + \frac{g_{bf}}{8\pi} + \frac{g_4}{8\pi} } }, \\
\tilde{v} &= \sqrt{ \left( \tilde{v}_F + \frac{g_4}{8\pi} \right)^2 - \left( \frac{g_{bf}}{8\pi} \right)^2 },
\end{align}
with $\tilde{v}_F = v_F \sqrt{1 - \left( \frac{\Delta_z}{\epsilon_F} \right)^2 }$ the Fermi velocity of the magnetized helical modes. The superconducting gap is effectively $\Delta = \Delta_s \sin(2 \vartheta_{k_F})$.
\begin{figure}[t!]
    \centering
    \includegraphics[width=9cm]{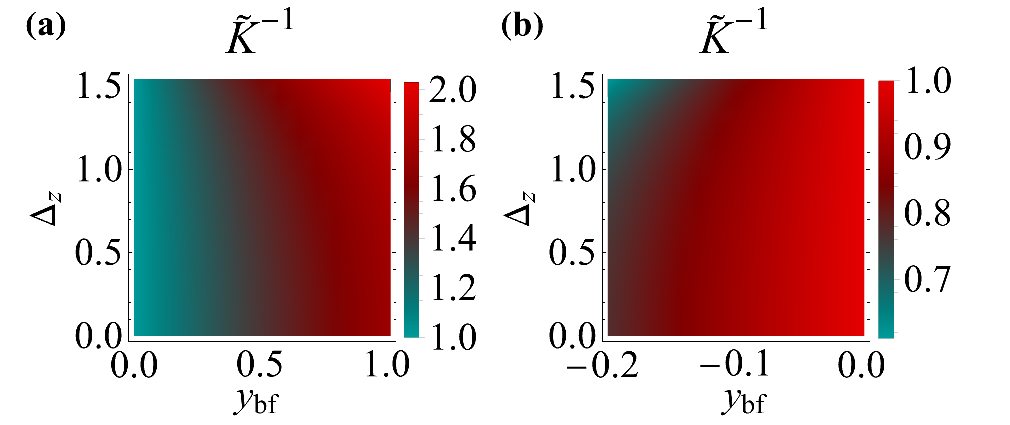}
    \caption{(Color online) Density plots showing the dependence of the inverse magnetized Luttinger parameter $\tilde{K}^{-1}$ on the interaction parameter $y_{bf}$ and the Zeeman gap $\Delta_z$ for (a) the repulsive and (b) attractive regimes.}
    \label{fig3}
\end{figure}

The corresponding RG flow equations are
\begin{align}
\frac{d \mathcal{Y}(l)}{d l} &= \left( 2 - \tilde{K}^{-1}(l) \right) \mathcal{Y}(l), \label{YRG} \\
\frac{d \tilde{K}^{-1}(l)}{d l} &= - \frac{1}{4} \mathcal{Y}^2(l) \tilde{K}^{-2}(l), \label{tildeKRG}
\end{align}
where $\mathcal{Y} = \frac{4 \Delta a_0}{\tilde{v}}$ is the dimensionless superconducting gap parameter. Equation~(\ref{YRG}) shows that $\mathcal{Y}$ becomes relevant for $\tilde{K} > \frac{1}{2}$, where $\tilde{K}$ is the Zeeman-renormalized Luttinger parameter. We further relate the magnetized Luttinger parameter $\tilde{K}$ to the interactions by introducing the magnetized-interaction parameter $\mathcal{Y}_{\vartheta_{k_F}}$:
\begin{equation}
\tilde{K}^{-1} = 2 + \mathcal{Y}_{\vartheta_{k_F}},
\label{KYtheta}
\end{equation}
where $\mathcal{Y}_{\vartheta_{k_F}} = y_{bf} - 1$ serves as an effective interaction parameter connecting the electron-electron interactions to $\tilde{K}$. Here, $y_{bf}$ is taken to equal $y_4$. Strong repulsive interactions drive the superconducting gap into the irrelevant regime. When $y_{bf}$ crosses the threshold of 1, and repulsive interactions weaken, the gap exits the marginal regime and becomes relevant. Without interactions, the system experiences a relevant gap. However, attractive interactions cause the superconducting gap to become more pronounced, pushing it deeper into the relevant regime as the attraction becomes stronger.

The behavior of the inverse Luttinger parameter $\tilde{K}^{-1}$ as a function of both the Zeeman gap $\Delta_z$ and the electron-electron interaction parameter $y_{bf}$ is illustrated in Fig. \ref{fig3}. In panel (a), which is for the repulsive interaction regime ($y_{bf} > 0$), $\tilde{K}^{-1}$ increases monotonically with $y_{bf}$. It also shows a modest increase with an increase in the Zeeman gap $\Delta_z$. So, increasing the Zeeman gap shifts the boundary for the relevance of the superconducting gap to smaller values of the repulsive interaction. Panel (b), which is for the attractive regime ($y_{bf} < 0$), shows that $\tilde{K}^{-1}$ is suppressed as the attraction (more negative $y_{bf}$) and $\Delta_z$ increase. So, in contrast to panel (a), an increase in the Zeeman gap enhances the strength of the superconducting gap, supporting the emergence of a robust pairing phase even for weaker attractive interactions.

\subsubsection{Zeeman-Affected Density and Pairing Correlation Functions}
\label{V-A-3}

In this subsection, we investigate how charge and spin density correlations, as well as superconducting pairing correlations, behave in the Zeeman-induced PMH state when a weak proximity pairing is added. Our goal is to determine, within this regime, which ordering tendency (CDW, SDW, or singlet/triplet superconductivity) remains dominant once superconductivity is turned on perturbatively, and how close to the phase boundary sizeable competing correlations persist. To this end, we calculate the corresponding correlation functions and analyze their modification by the combined Zeeman–pairing gap, which enters as a sine-Gordon term arising from the superconducting coupling to the magnetized helical wire.

Equations (\ref{CDW-op-equ})–(\ref{TS-op-equ}) yield the correlation operators for the PMH edge as
\begin{align}
    \mathcal{O}_{CDW}(x,\tau)&=-\frac{1}{\sqrt{4\pi}}\partial_x\Phi(x,\tau)\nonumber\\&-\frac{\sin(2\vartheta_{k_F)}}{\pi a_{0}}\cos(\sqrt{4\pi}\Phi(x,\tau)+2k_{F}x),
 \label{CDWOp1}
 \\\mathcal{O}^{x}_{SDW}(x,\tau)&=-\frac{\cos(2\vartheta_{k_F})}{\sqrt{4\pi}}\partial_x\Theta(x,\tau),
  \label{xSDWOp1}
  \\
 \mathcal{O}^{y}_{SDW}(x,\tau)&=\frac{\sin(2\vartheta_{k_F})}{\sqrt{4\pi}}\partial_x\Phi(x,\tau)\nonumber\\&+\frac{1}{\pi a_{0}}\cos(\sqrt{4\pi}\Phi(x,\tau)+2k_{F}x),
   \label{ySDWOp1}\\
 \mathcal{O}^{z}_{SDW}(x,\tau)&=-\frac{\cos(2\vartheta_{k_F})}{\pi a_{0}}\sin(\sqrt{4\pi}\Phi(x,\tau)+2k_{F}x),
 \label{zSDWOp1}\\\mathcal{O}_{SS}(x,\tau)&=\frac{\cos(2\vartheta_{k_F})}{2\pi a_0}e^{-\sqrt{4\pi}\Theta(x,\tau)},\label{ss-pair-op}\\
 \mathcal{O}^{x}_{TS}(x,\tau)&=-\frac{e^{-\sqrt{4\pi}\Theta(x,\tau)}}{2\pi a_0},\label{x-pair-op}\\\mathcal{O}^{z}_{TS}(x,\tau)&=-\frac{i\sin(2\vartheta_{k_F})}{2\pi a_0}e^{-\sqrt{4\pi}\Theta(x,\tau)}.
 \label{z-pair-op}
\end{align}
Inserting Eqs.~(\ref{CDWOp1})–(\ref{z-pair-op}) into the correlation function definitions (\ref{RCDW})–(\ref{RTS}), the magnetized correlations with respect to $\mathcal{H}_{PMH}$ are obtained as
\begin{align}
\mathcal{R}_{CDW}(r)&=\frac{1}{4\pi^2}\tilde{K}\tilde{F}_2(r)\\&+\frac{\sin^2(2\vartheta_{k_F})\cos(2k_Fx)}{2(\pi a_0)^2}e^{-2\tilde{K}\tilde{F}_3(r)},
\label{CDWsemihel}\\
\mathcal{R}^{xx}_{SDW}(r)&=-\frac{\cos^2(2\vartheta_{k_F})}{4\pi^{2}\tilde{K}}\tilde{F}_2(r),\label{xSDWsemihel}\\
\mathcal{R}^{yy}_{SDW}(r)&=-\frac{\sin^2(2\vartheta_{k_F})}{4\pi^{2}}\tilde{K}\tilde{F}_2(r)+\frac{\cos(2k_F x)}{2(\pi a_0)^2}e^{-2\tilde{K}\tilde{F}_{3}(r)},\label{ySDWsemihel}\\
\mathcal{R}^{zz}_{SDW}(r)&=\frac{\cos^2(2\vartheta_{k_F})\cos(2k_F x)}{2(\pi a_0)^2}e^{-2\tilde{K}\tilde{F}_{3}(r)},\label{zSDWsemihel}\\
\mathcal{R}^{yz}_{SDW}(r)&=-\mathcal{R}^{zy}_{SDW}(r)\nonumber\\&=-\frac{\cos^2(2\vartheta_{k_F})\cos(2k_F x)}{2(\pi a_0)^2}e^{-2\tilde{K}\tilde{F}_{3}(r)},
\label{yzSDWsemihel}\\\mathcal{R}_{SS}(r)&=\frac{\cos^2(2\vartheta_{k_F})}{(2\pi a_0)^2}e^{-2\tilde{K}^{-1}\tilde{F}_{3}(r)},\label{ss-semihel}\\\mathcal{R}^{x}_{TS}(r)&=\frac{1}{(2\pi a_0)^2}e^{-2\tilde{K}^{-1}\tilde{F}_{3}(r)},\label{x-TS-semihel}\\\mathcal{R}^{z}_{TS}(r)&=\frac{\sin^2(2\vartheta_{k_F})}{(2\pi a_0)^2}e^{-2\tilde{K}^{-1}\tilde{F}_{3}(r)},\label{z-TS-semihel}
\end{align}
where the functions $\tilde{F}_2(r)$ and $\tilde{F}_3(r)$ are defined as
\begin{align}
\tilde{F}_2(r)&=\frac{(\tilde{v}\tau sign(\tau)+a_0)^{2}-x^2}{2((\tilde{v}\tau sign(\tau)+a_0)^{2}+x^2)^2},
\label{F2}\\
\tilde{F}_{3}(r)&=\frac{1}{2}log(\frac{(\tilde{v}\tau sign(\tau)+a_0)^{2}+x^2}{a_0^{2}}).
\label{F3}
\end{align}
We note that the effect of the Zeeman field is encoded in the Luttinger parameters $\tilde{K}$ and $\tilde{v}$ as well as in the spin mixing angle $\vartheta_{k_F}$.

SDW states naturally emerge in low-dimensional electron systems due to Fermi surface nesting and strong electron-electron correlations. The x-phase, in particular, is characterized by a helical spin modulation resulting from spontaneous breaking of spin-rotational symmetry. Application of an external Zeeman field destabilizes this helical spin configuration by explicitly breaking helicity conservation. This symmetry breaking manifests as deformation or suppression of the spin texture, depending sensitively on the magnitude and orientation of the field.

\begin{figure*}[t!]
    \centering
    \includegraphics[width=15cm]{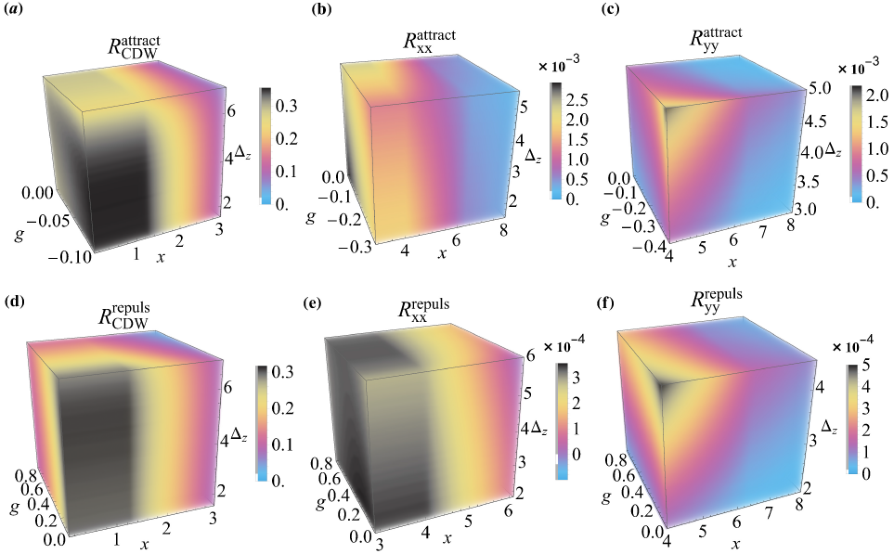}
    \caption{(Color online) Short‐range correlation amplitudes $R(x;g,\Delta_z)$ in the partially mixed helical wire, plotted versus interaction strength $g$, separation $x$, and Zeeman gap $\Delta_z$. Panels (a)–(c) show the attractive regime ($g<0$): (a) charge‐density‐wave $R^{\rm attract}_{\rm CDW}$, (b) longitudinal SDW $R^{\rm attract}_{xx}$, (c) transverse SDW $R^{\rm attract}_{yy}$. Panels (d)–(f) show the repulsive regime ($g>0$): (d) charge‐density‐wave $R^{\rm repuls}_{\rm CDW}$, (e) longitudinal SDW $R^{\rm repuls}_{xx}$, (f) transverse SDW $R^{\rm repuls}_{yy}$. The color scale indicates the magnitude of each correlation function.}
   \label{shortrange}
\end{figure*}

Figure \ref{shortrange} shows the short-range behavior of the corrected correlation functions for CDW and SDW orders in the PMH regime versus the interaction strength $g$, distance $x$, and Zeeman gap $\Delta_z$. The first row, panels (a)-(c), corresponds to attractive interactions, $g<0$; the second row, panels (d)-(f), to repulsive interactions, $g>0$.

Panels (a) and (d) present the short-range CDW amplitudes for attractive and repulsive interactions, respectively. The CDW amplitude exhibits a pronounced peak at large negative $g$ and small separations, signifying robust local charge order when interactions are strongly attractive and the Zeeman gap is weak. The CDW correlation function decreases with increasing Zeeman field, and this suppression is further enhanced when the Zeeman term becomes more relevant due to electron-electron interactions. Under repulsive interactions, the maximum amplitude of the CDW shifts toward small positive values of $g$. In this regime as well, increasing the Zeeman field leads to a reduction in the CDW correlation. Stronger repulsion and larger Zeeman gap further reinforce this suppression. In both cases, increasing the separation suppresses the CDW rapidly, establishing that the enhancement of short-range CDW correlations is confined to a strong coupling and a minimal Zeeman field.

Panels (b) and (e) show the corresponding short-range amplitudes for the SDW$_{xx}$ in attractive and repulsive regimes, respectively. Throughout the parameter space, attractive and repulsive interactions result in SDW$_{xx}$ amplitudes that are smaller, by more than three orders of magnitude, than those of the CDW channel. Interestingly, unlike in the CDW case, a strong peak takes place at a small negative interaction corresponding to a deeper Zeeman gap. This peak diminishes as the system goes toward stronger attractive interactions with the weaker Zeeman gap. Moreover, SDW$_{xx}$ suppresses with increasing Zeeman strength in both interacting regimes. This non-monotonic behavior implies the subtle competition between external magnetic field and the electron-electron interaction. However, similar to the CDW case, in both attractive and repulsive regimes the amplitudes remain negligible and decrease rapidly with increasing distance.

Panels (c) and (f) display the behavior of the SDW$_{yy}$ correlation in attractive and repulsive regimes, respectively. The SDW$_{yy}$ channel exhibits negligible amplitudes seen in the longitudinal spin sector, regardless of whether interactions are attractive or repulsive. Here, a large enhancement is observed at both large attractive interaction and large Zeeman field. This is in contrast to usual cases where a strong Zeeman field tends to break spin rotational symmetry and suppress transverse spin order. However, in the present case, the renormalization of the Zeeman coupling due to interactions can reinforce this component of the SDW. Consequently, even at fixed Zeeman strength, stronger attractive interactions enhance effective magnetic fields felt by spins, potentially stabilizing or restoring long-range correlations weakened by external perturbations. Furthermore, in all cases, the amplitude is quickly diminished as the distance is increased. Thus, local transverse spin-density fluctuations remain well below the leading CDW response in both interaction regimes and cannot develop into dominant short-range order.

\begin{figure*}[t!]
    \centering
    \includegraphics[width=18cm]{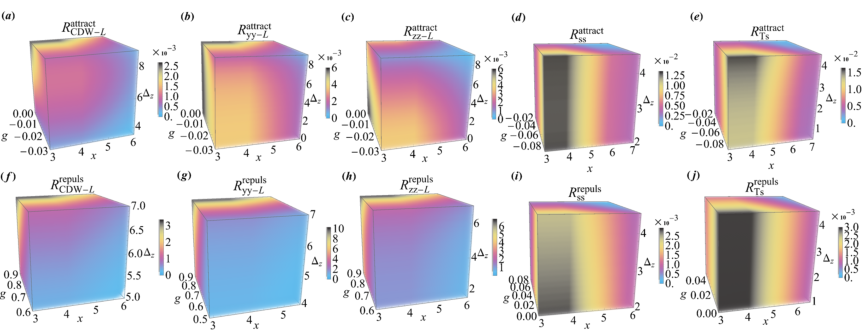}
   \caption{(Color online) Long-range density‐wave and pairing correlation amplitudes $R(x;g,\Delta_z)$ in the partially mixed helical (PMH) wire, shown as functions of distance $x$, interaction strength $g$, and Zeeman gap $\Delta_z$. Panels (a)–(e) correspond to attractive interactions ($g<0$), displaying (a) $R^{\rm attract}_{\rm CDW\text{-}L}$, (b) $R^{\rm attract}_{yy\text{-}L}$, (c) $R^{\rm attract}_{zz\text{-}L}$, (d) $R^{\rm attract}_{\rm SS}$ (singlet pairing), and (e) $R^{\rm attract}_{\rm TS}$ (triplet-$x$ pairing). Panels (f)–(j) correspond to repulsive interactions ($g>0$), displaying (f) $R^{\rm repuls}_{\rm CDW\text{-}L}$, (g) $R^{\rm repuls}_{yy\text{-}L}$, (h) $R^{\rm repuls}_{zz\text{-}L}$, (i) $R^{\rm repuls}_{\rm SS}$, and (j) $R^{\rm repuls}_{\rm TS}$. The color scale at each panel indicates the magnitude of the correlation function.
}\label{longrange}
\end{figure*}

Figure \ref{longrange} provides an analysis of the long-range behavior of the density wave and superconducting pairing correlations within the PMH state. The figure shows the correlation amplitudes as a function of spatial separation $x$, interaction strength $g$, and Zeeman gap $\Delta_z$. The results are organized into two main regimes: attractive interactions (panels (a)-(e)) and repulsive interactions (panels (f)-(j)).

Panels (a) and (f) show the behavior of the long-range CDW correlations for attractive and repulsive interactions, respectively.  For repulsive interactions, the CDW amplitude is most prominent, reaching a maximum at strong coupling and highest $\Delta_z$. It decreases rapidly as $\Delta_z$ decreases. In the attractive case, the CDW remains vanishingly small over a broad range of parameters. But, its enhancement for strong Zeeman gaps at weak coupling is less pronounced.

Panels (b) and (g) show the SDW$_{yy}$ component. For attractive interactions (panel b), the SDW$_{yy}$ is highly suppressed, maintaining only negligible amplitudes at weak coupling. Such amplitudes are also less sensitive to $\Delta_z$, indicating that Zeeman fields cannot induce substantial spin ordering in this regime. In contrast, under repulsion (panel g), the SDW$_{yy}$ correlation exhibits a large amplitude at strong couplings. Moreover, increasing the Zeeman field strengthens the long-range SDW$_{yy}$ correlations and promotes a dominant ordering tendency, possibly leading to a leading spin phase.

Panels (c) and (h) present the SDW$_{zz}$ correlations. Under attractive interactions (panel c), the SDW$_{zz}$ remains almost suppressed, revealing substantial suppression by magnetic effect and attractive correlation. It generally remains subdominant throughout the parameter space. For repulsive interactions (panel h), interestingly, this component shows a similar profile compared with the transverse SDW$_{yy}$, with a maximum appearing at larger $\Delta_z$ and strong repulsion. Although its amplitude remains smaller compared to the SDW in the $y$-direction.

Panels (d) and (i), show the singlet superconducting (SS) correlation amplitude. With attractive interactions (panel d), the SS channel is the strongest in the regime of strong attraction and for any given Zeeman field. This implies the stability of conventional superconducting pairing. The amplitude unchanges as $\Delta_z$ changes. Under repulsive interactions (panel i), the singlet correlation drops and the increase of the Zeeman field reduces this correlation.

Panels (e) and (j) illustrate the triplet-x superconducting component. Notably, in the attractive regime (panel e), the triplet correlation is strongly enhanced by increasing the Zeeman field, yielding a distinct ridge in parameter space. In this regime, triplet-x component is a dominant phase. This is consistent with the theoretical expectation that attractive interactions and finite Zeeman splitting favor the stabilization of triplet pairing. However, under repulsion (panel j), the triplet amplitude is consistently small across all parameters, indicating that triplet pairing does not emerge in this regime. It also is insensitive to the Zeeman field in this regime. Remarkably, the predominance of triplet-$x$ pairing also suggests that the Zeeman field may induce a topologically nontrivial superconducting phase characterized by unconventional triplet pairing.

It is important to note that in the strong-coupling regime of Zeeman (see Section \ref{V-A-1}), these pairing excitations are absent. So, only the $x$- and $y$-phases survive, with the $y$-phase dominating. In contrast, at finite Zeeman fields, we observe an enhancement in not only the CDW but also the $z$ and mixed components of the SDW correlations, as well as the singlet and triplet pairing correlations. Even more interestingly, at the relevancy boundary of the Zeeman gap, corresponding to the weak Zeeman case discussed in Section \ref{V-B-2}, the Zeeman-induced logarithmic correction to the correlations notably suppresses the $y$-phase while enhancing the pairing correlations.

\subsubsection{Logarithmic Corrections Arising from Pairing}
\label{V-A-4}

In this section, we calculate the logarithmic corrections to the main correlation functions arising from the marginal pairing gap. We rewrite the RG flow equations (\ref{YRG}) and (\ref{tildeKRG}) in terms of the magnetized-interaction parameter $\mathcal{Y}_{\vartheta_{k_F}}$ as
\begin{align}
\frac{d \mathcal{Y}(l)}{dl} &= - \mathcal{Y}_{\vartheta_{k_F}} \mathcal{Y}(l), \label{YRGint} \\
\frac{d \mathcal{Y}_{\vartheta_{k_F}}(l)}{dl} &= - \mathcal{Y}^2(l).
\label{tildeKRGint}
\end{align}
The above equations are obtained by expansion up to second order in the coupling constants $\mathcal{Y}$ and $\mathcal{Y}_{\vartheta_{k_F}}$. To find the RG flow trajectories, we multiply Eq.~(\ref{YRGint}) by $\mathcal{Y}(l)$ and Eq.~(\ref{tildeKRGint}) by $\mathcal{Y}_{\vartheta_{k_F}}(l)$, leading to the conserved quantity
\begin{equation}
A_z^2 = \mathcal{Y}_{\vartheta_{k_F}}^2 - \mathcal{Y}^2.
\end{equation}
Here, $A_z=0$ corresponds to the boundary between relevant and irrelevant gap regimes, while $A_z \neq 0$ characterizes exact regimes. Inserting this constant of motion into Eqs.~(\ref{YRG}) and (\ref{tildeKRG}) and integrating yields

\begin{align}
\mathcal{Y}_{\vartheta_{k_F}}(l) &= \frac{A_z}{\tanh\left( A_z l + a \tanh \frac{A_z}{\mathcal{Y}_{\vartheta_{k_F}}(0)} \right)}, \label{soluyint} \\
\mathcal{Y}(l) &= \frac{A_z}{\sinh\left( A_z l + a \tanh \frac{A_z}{\mathcal{Y}_{\vartheta_{k_F}}(0)} \right)}. \label{soluy}
\end{align}

Using these solutions, we evaluate the pairing-induced corrections to the density correlation functions as

\begin{align}
\mathcal{R}^{\text{log-pair}}_{CDW}(r) &= \frac{\sin^2(2 \vartheta_{k_F}) \cos(2 k_F x)}{2 (\pi a_0)^2} \frac{a_0}{r} L_1, \label{CDW-pair-perturb} \\
\mathcal{R}^{yy, \text{log-pair}}_{SDW}(r) &= \frac{\cos(2 k_F x)}{2 (\pi a_0)^2} \frac{a_0}{r} L_1, \label{ySDW-pair-perturb} \\
\mathcal{R}^{zz, \text{log-pair}}_{SDW}(r) &= \frac{\cos^2(2 \vartheta_{k_F}) \cos(2 k_F x)}{2 (\pi a_0)^2} \frac{a_0}{r} L_1, \label{zSDW-pair-perturb} \\
\mathcal{R}^{yz, \text{log-pair}}_{SDW}(r) &= - \mathcal{R}^{zy, \text{log-pair}}_{SDW}(r)\\& = - \frac{\cos^2(2 \vartheta_{k_F}) \cos(2 k_F x)}{2 (\pi a_0)^2} \frac{a_0}{r} L_1, \label{yzSDW-pair-perturb}
\end{align}
where
\begin{equation}
L_1 = \exp \left[ \frac{1}{2} \int_0^{l_r} \mathcal{Y}_{\vartheta_{k_F}}(l) dl \right].
\end{equation}
The corresponding correction for pairing correlations (except for $\mathcal{R}^{z}_{TS}$, because it can be ignored) reads
\begin{align}
\mathcal{R}^{\text{log-pair}}_{SS}(r) &= \frac{\cos^2(2 \vartheta_{k_F})}{(2 \pi a_0)^2} \left( \frac{a_0}{r} \right)^4 L_2, \label{ss-pair-perturb} \\
\mathcal{R}^{x, \text{log-pair}}_{TS}(r) &= \frac{1}{(2\pi a_0)^2} \left( \frac{a_0}{r} \right)^4 L_2, \label{x-TS-pair-perturb}
\end{align}
with
\begin{equation}
L_2 = \exp \left[- 2 \int_0^{l_r} \mathcal{Y}_{\vartheta_{k_F}}(l) dl \right].
\end{equation}

At the marginal boundary $\mathcal{Y}_{\vartheta_{k_F}} = \mathcal{Y}$, the solutions simplify to
\begin{equation}
\mathcal{Y}(l) = \mathcal{Y}_{\vartheta_{k_F}}(l) = \frac{\mathcal{Y}_{\vartheta_{k_F}}(0)}{1 + \mathcal{Y}_{\vartheta_{k_F}}(0) l},
\end{equation}
and in the limit $r \to \infty $,
\begin{align}
L_1 &= \mathcal{Y}_{\vartheta_{k_F}}^{1/2} \log^{1/2} \left( \frac{r}{a_0} \right), \label{L_1} \\
L_2 &= \mathcal{Y}_{\vartheta_{k_F}}^{-2} \log^{-2} \left( \frac{r}{a_0} \right). \label{L_2}
\end{align}
Since the marginal boundary of superconductivity shifts towards weaker repulsive interactions with increasing Zeeman strength, these logarithmic corrections appear in the correlations for weaker repulsion. According to Eqs. (\ref{CDW-pair-perturb})–(\ref{yzSDW-pair-perturb}), all density wave correlations exhibit the same power-law decay characteristic of the topological insulator edge, with superconductivity introducing an incremental logarithmic correction. In contrast, pairing correlations, which decay faster, receive decreasing logarithmic corrections from superconductivity. This suggests that in a 1D topological superconductor with relevant Zeeman and marginal superconductivity, density wave excitations dominate, and superconductivity enhances their strength.

In the irrelevant regime ($ A_z l \gg 1 $), using the general solutions (\ref{soluyint}) and (\ref{soluy}) to integrate $ L_1 $ and $L_2 $, we find

\begin{align}
L_1 &\simeq \frac{1}{\sqrt{2}} \left( \frac{a_0}{r} \right)^{- \frac{A_z}{2}} \left( 1 + \frac{\mathcal{Y}_{\vartheta_{k_F}}}{A_z} \right)^{1/2}, \label{L_1irelev} \\
L_2 &\simeq 4 \left( \frac{a_0}{r} \right)^{2 A_z} \left( 1 + \frac{\mathcal{Y}_{\vartheta_{k_F}}}{A_z} \right)^{-2}. \label{L_2irrelev}
\end{align}
No logarithmic corrections appear, and the results resemble the non-perturbative case but with additional coefficients depending on the magnetized interaction parameter and the renormalized Luttinger parameter:

\begin{align}
\tilde{K}^* &= \frac{1}{2} \left( 1 - \frac{A_z}{2} \right), \quad \text{(density)}, \\
\tilde{K}^{*-1} &= 2 + A_z, \quad \text{(pairing)}.
\end{align}
Here, $l = \log \frac{r}{a_0}$ defines the characteristic length scale separating regimes dominated by logarithmic corrections and non-perturbative renormalization of parameters.

\subsubsection{Pairing-Induced Correction of Short-Range Correlations in the PMH Edge}
\label{V-A-5}

In this subsection, we calculate corrections to the short-range components of the density correlation functions arising from perturbation theory up to second order in the superconducting gap parameter. This approach allows us to study the effects of an irrelevant gap (in the RG sense) on the residual parts of the correlation functions. Ultimately, the full correction to each correlation function is expressed as a superposition of logarithmic corrections (treated in the previous section) and these residual corrections.

Since the logarithmic corrections were derived in real space and time, we combine those results here with Fourier space calculations. Corrections to the first term of the CDW operator (Eq. (\ref{CDWOp1})) and the SDW operator in the $x$ direction (Eq. (\ref{xSDWOp1})) are defined as
 \begin{align}
     &\mathcal{C}_{CDW}(r)=\frac{\mathcal{Y}^2(l)}{128\pi^3 a_0^4}\sum_{\epsilon=\pm1}\nonumber\\&\times\langle T_{\tau}\partial_{x}\Phi(r)\partial_{x}\Phi(0)e^{i\sqrt{4\pi}\epsilon \Theta(r_3)}e^{-i\sqrt{4\pi}\epsilon \Theta(r_4)}\rangle_{\mathcal{H}_{PMH}},
 \end{align}
and 
  \begin{align}
     &\mathcal{C}^{xx}_{SDW}(r)=\frac{\mathcal{Y}^2(l)\cos^2(2\vartheta_{k_F})}{128\pi^3 a_0^4}\sum_{\epsilon=\pm1}\nonumber\\&\times\langle T_{\tau}\partial_{x}\Theta(r)\partial_{x}\Theta(0)e^{i\sqrt{4\pi}\epsilon \Theta(r_3)}e^{-i\sqrt{4\pi}\epsilon \Theta(r_4)}\rangle_{\mathcal{H}_{PMH}},
 \end{align}
The expected values in these expressions can be evaluated by known correlators of the form
 \begin{equation}
     \sum_{\epsilon=\pm1}\langle T_{\tau}e^{i\lambda\partial_{x}\Phi(r)}e^{i\mu\partial_{x}\Phi(0)}e^{i\sqrt{4\pi}\epsilon \Theta(r_3)}e^{-i\sqrt{4\pi}\epsilon \Theta(r_4)}\rangle_{\mathcal{H}_{PMH}},
 \end{equation}
and
 \begin{equation}
     \sum_{\epsilon=\pm1}\langle T_{\tau}e^{i\lambda\partial_{x}\Theta(r)}e^{i\mu\partial_{x}\Theta(0)}e^{i\sqrt{4\pi}\epsilon \Theta(r_3)}e^{-i\sqrt{4\pi}\epsilon \Theta(r_4)}\rangle_{\mathcal{H}_{PMH}}.
 \end{equation}
As shown in Refs.~\cite{Bakhshipour2024,Citro2020}, after evaluation the correction terms assume the well-known form involving Hartree and Fock averages:
  \begin{widetext}
      \begin{equation}
     \mathcal{C}_{CDW}(r)=\frac{\mathcal{Y}^2(l)}{8\pi^2 a_0^4}\int d^2r_3 d^2r_4 e^{-2\pi \langle T_{\tau}[\Theta(r_3)-\Theta(r_4)]^2\rangle}\\\langle T_{\tau}[\partial_x \Phi(r)\Theta(r_3)]\rangle\langle T_{\tau}[\partial_x \Phi(0)\Theta(r_3)-\partial_x \Phi(0)\Theta(r_4)]\rangle,
     \label{1correlationresidual}
 \end{equation}
 \begin{equation}
      \mathcal{C}^{xx}_{SDW}(r)=\frac{\mathcal{Y}^2(l)\cos^2(2\vartheta_{k_F})}{8\pi^2 a_0^4}\int d^2r_3 d^2r_4 e^{-2\pi \langle T_{\tau}[\Theta(r_3)-\Theta(r_4)]^2\rangle}\\\langle T_{\tau}[\partial_x \Theta(r)\Theta(r_3)]\rangle\langle T_{\tau}[\partial_x \Theta(0)\Theta(r_3)-\partial_x \Theta(0)\Theta(r_4)]\rangle.
      \label{2correlationresidual}
 \end{equation}
  \end{widetext}
It can be shown that the Hartree term does not contribute to the correction when $\tilde{K}^{-1} > 2$, whereas the Fock term exhibits physically relevant behavior in this regime. Specifically, the short-range correlation corrections in the frequency space behave as
\begin{align}
&\mathcal{R}^{short-range}_{CDW}(\omega)\approx\omega +\mathcal{Y}^2\omega^{2\tilde{K}^{-1}-3},
\\
&\mathcal{R}^{xx, short-range}_{SDW}(\omega)\approx\omega+\mathcal{Y}^2\cos^2(2\vartheta_{k_F})\omega^{2(\tilde{K}^{-1}-2)},
\\
&\mathcal{R}^{yy, short-range}_{SDW}(\omega)\approx\omega+\mathcal{Y}^2\sin^2(2\vartheta_{k_F})\omega^{2\tilde{K}^{-1}-3}.
\end{align}
Notably, the charge density phase at $\tilde{K}^{-1} > 2$ is unaffected by superconducting corrections, so the correction term becomes negligible as $\omega \to 0$. For $\tilde{K}^{-1} < 2$, corrections become significant. The SDW phase in the $x$ direction undergoes superconducting corrections for $\tilde{K}^{-1} < 5/2$, indicating competition between helicity-aligned SDW short-range order and long-range density wave excitations under pairing. The $y$-component of SDW behaves similarly to CDW but is weighted by $\sin^2(2\vartheta_{k_F})$.

In topological superconducting PMH states with relevant Zeeman and marginal superconductivity, long-range density wave excitations under repulsive interactions are enhanced by superconductivity. For the short-range $x$-component SDW, the relevancy boundary shifts toward repulsive interactions beyond $1$, implying that stronger Zeeman fields are needed to suppress superconductivity in this channel, consistent with discussions in Section~\ref{V-A-2}.

\subsection{Pairing-induced superhelical state perturbed by Zeeman effect} 
\label{V-B}

In this part, we consider the pairing gap to be relevant. This means that the Zeeman gap dwells near the boundary of its weak regime. We then calculate the logarithmic correction of this marginal Zeeman gap to the superhelical wire. Here, the field operators are taken with $\vartheta_{k_F} = 0$. The effective superconducting helical state Hamiltonian $\mathcal{H} = \mathcal{H}_{s-hel} + \mathcal{H}_z$ reads
\begin{equation}
\begin{split}
\mathcal{H} = \frac{\bar{v}}{2} \left[ \frac{1}{\bar{K}} (\partial_x \Phi)^2 + \bar{K} (\partial_x \Theta)^2 \right] 
+ \frac{\tilde{\Delta}_z}{2 \pi a_0} e^{i \sqrt{4\pi} \tilde{\Phi}} + \mathrm{h.c.}.
\label{Hamfirst1}
\end{split}
\end{equation}
The superconducting Luttinger liquid parameter $\bar{K}$ and velocity $\bar{v}$ are given by
\begin{equation}
\bar{K} = \sqrt{\frac{\bar{v}_F - \frac{g_{bf}}{8\pi} + \frac{g_4}{8\pi}}{\bar{v}_F + \frac{g_{bf}}{8\pi} + \frac{g_4}{8\pi}}},
\label{Luttingerparameter}
\end{equation}
\begin{equation}
\bar{v} = \sqrt{\left(\bar{v}_F + \frac{g_4}{8\pi}\right)^2 - \left(\frac{g_{bf}}{8\pi}\right)^2}.
\label{Luttingervelocity}
\end{equation}
Here, $\bar{v}_F = v_F \sqrt{1 - \left( \frac{\Delta_s}{\epsilon_F} \right)^2 }$ is the effective Fermi velocity. Note that in Eq. (\ref{Hamfirst1}) the pairing term is not written explicitly because we work in the regime where superconducting pairing is a relevant operator and has already flowed to strong coupling. In this limit, the pairing cosine pins the dual
field $\Theta$ and generates a finite gap. Therefore, its effect is absorbed into the renormalized Fermi velocity $\bar v_F$ and Luttinger parameter $\bar K$ entering Eqs. (\ref{Luttingerparameter})--(\ref{Luttingervelocity}). Equation (\ref{Hamfirst1}) should therefore be understood as an effective low-energy description of the superhelical phase, in which the spectrum is gapped by pairing while the Zeeman term provides the remaining marginal perturbation whose RG flow we analyze below.

The RG flow equations for the Zeeman coupling and the superconducting Luttinger parameter can be obtained as
\begin{align}
\frac{d \bar{K}(l)}{dl} &= -\frac{1}{4} \mathcal{Y}_z^2(l) \bar{K}^2(l), 
\label{KRG-s-eff} \\
\frac{d \mathcal{Y}_z(l)}{dl} &= \left(2 - \bar{K}(l)\right) \mathcal{Y}_z(l),
\label{YzRG-s-eff}
\end{align}
where $\mathcal{Y}_z = \frac{\sqrt{2} \Delta_z a_0}{\bar{v}}$ is the dimensionless Zeeman gap coupling. According to Eq.~(\ref{YzRG-s-eff}), the Zeeman coupling $\mathcal{Y}_z$ becomes relevant when $\bar{K} < 2$. It is observed that, unlike Eq. (\ref{YzRG}), the scaling of the Zeeman coupling coefficient here depends on the pairing-renormalized Luttinger parameter, which itself is influenced not only by the electron-electron interactions but also by the superconducting gap.

\begin{figure}[t!]
    \centering
    \includegraphics[width=9cm]{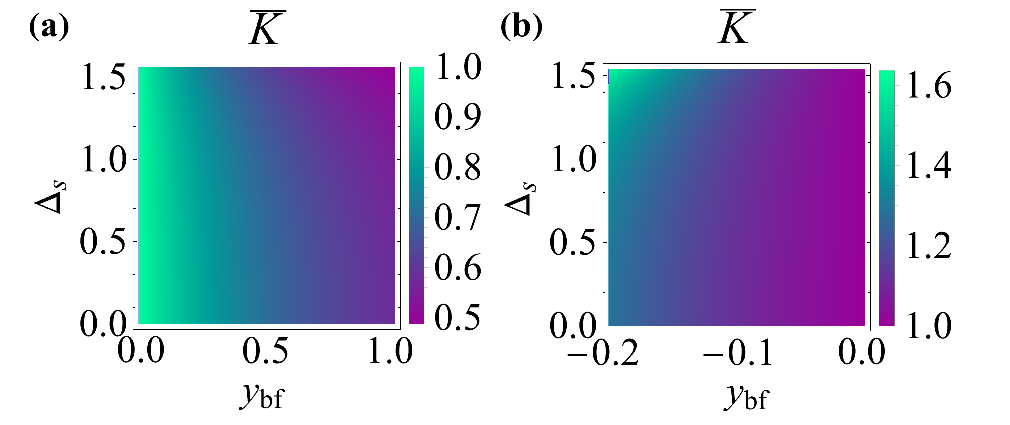}
    \caption{
        (Color online) Density plot showing the dependence of the superconducting  Luttinger parameter $\bar{K}$ on the pairing gap $\Delta_s$ and the interaction parameter $y_{bf}$ for (a) the repulsive and (b) the attractive interactions.
    }
    \label{fig4}
\end{figure}

Moreover, we find a relation between the Luttinger parameter $\bar{K}$ and the interactions by defining the pairing-gap-tuning interaction parameter $\bar{\mathcal{Y}}_{int}$ as
\begin{equation}
\bar{K} = 2 - \bar{\mathcal{Y}}_{int},
\label{KYint-s}
\end{equation}
where $\bar{\mathcal{Y}}_{int} = 1 + y_{bf}$. Here, in the absence of a pairing gap, as discussed in Section \ref{III}, the Zeeman term becomes relevant for both weak attractive and repulsive interactions with strengths less than 1. However, once superconducting pairing sets in, these boundaries shift. 

In Fig. \ref{fig4}, the dependence of the superconductivity-renormalized Luttinger parameter $\bar{K}$ on the pairing gap $\Delta_s$ and the interaction parameter $y_{bf}$ is shown for the repulsive regime, $y_{bf} > 0$, (panel a) and the attractive regime, $y_{bf} < 0$, (panel b). 
As shown in Fig. \ref{fig4}(a),  increasing the repulsive interaction strength reduces $\bar{K}$ so that it takes values below unity. This indicates that repulsion suppresses $\bar{K}$, resulting in stabilization of the Zeeman gap in the system. Increasing $\Delta_s$, slightly reduces $\bar{K}$. In contrast, Fig. \ref{fig4}(b) illustrates the influence of attractive interactions on the Zeeman gap. As the magnitude of the attractive interaction increases, the value of $\bar{K}$ increases, so that it reaches values above unity. Similarly, a larger $\Delta_s$ increases the value of $\bar{K}$. Therefore, in this regime, the system becomes unstable toward the formation of superconductivity over magnetism.

\subsubsection{Strong Pairing with Weak Zeeman Field}
\label{V-B-1}

A comprehensive view of the phase diagram requires exploring the strong-coupling regime of pairing. Notably, in many systems, such as the Kitaev model, this regime naturally leads the system into a trivial topological phase \cite{Kitaev2001,Shi2023,Cinnirella2025}. In proximity-induced systems, such as nanowires or two-dimensional heterostructures, a sufficiently strong parent superconductor may induce strong pairing in the wire. Investigating this regime becomes particularly important in the presence of magnetic fields or disorder.

According to our analysis, in the strong pairing limit, the bosonic field $\Theta$ becomes ordered. As a result, all power-law correlations involving the dual field $\Phi$ are suppressed. The CDW operator, which involves the gradient of the bosonic field $\Phi$, benefits from its strong fluctuations and maintains its Fermi-like behavior with parameters renormalized by superconductivity. The SDW$_{xx}$ vanishes, while the singlet and triplet-$x$ superconducting correlations still exhibit power-law decay, though slower than before, eventually saturating to a finite value.

Topologically, in contrast to the weak pairing regime, where the system enters a topological superconducting phase with Majorana edge states and dominant singlet correlations, in the strong pairing regime, the system typically flows into a trivial topological phase. Due to strong pairing, the fermions become locally bound and no edge states emerge, leading to dominant short-range and local correlations. Interestingly, in our study, despite the strong coupling, the singlet and triplet-$x$ superconducting correlations remain the dominant ones.


\subsubsection{Logarithmic Corrections Arising from Zeeman}
\label{V-B-2}

We use the pairing-gap-dependent interaction parameter, denoted as $\bar{\mathcal{Y}}_{int}$, and rewrite the RG flow equations for the Zeeman coupling $\mathcal{Y}_z$ and the interaction parameter by eliminating the Luttinger parameter $\bar{K}$. Keeping terms up to second order, the flow equations read:

\begin{align}
\frac{d\bar{\mathcal{Y}}_{int}(l)}{dl}&=\mathcal{Y}_z^2(l)
\label{barKRGint},
\\\frac{d\mathcal{Y}_z(l)}{dl}&=\bar{\mathcal{Y}}_{int}\mathcal{Y}_z(l).
\label{YzRGint}
\end{align}
From these equations, we can obtain a constant motion defining the marginal Zeeman boundary in the presence of superconductivity:
\begin{equation}
    A_{s}^2=\bar{\mathcal{Y}}^2_{int}-\mathcal{Y}^2.
\end{equation}

At this boundary (the separatrix), logarithmic corrections to the density and pairing correlation functions arise. The correction to the SDW correlation along the $y$ obtains
\begin{align}
\mathcal{R}^{yy,log,Ze}_{SDW}(r)=\frac{\cos(\alpha x)}{2(\pi a_0)^2}(\frac{a_0}{r})^4(\bar{\mathcal{Y}}_{int}(l))^{-2}log^{-2}(\frac{r}{a_0})\label{ySDWnonperturb}.
\end{align}
The $zz$ component and mixed components $\mathcal{R}^{yz,\mathrm{log},\mathrm{Ze}}{SDW}(r) = -\mathcal{R}^{zy,\mathrm{log},\mathrm{Ze}}{SDW}(r)$ behave similarly but with different coefficients. The pairing correlations along the separatrix behave as
\begin{align}
    &\mathcal{R}^{log,Ze}_{SS}(r)=\mathcal{R}^{xx,log,Ze}_{TS}(r)\nonumber\\&=-\frac{1}{(2\pi a_0)^2}(\frac{a_0}{r})^{2-\frac{3}{2}\mathcal{Y}_{eff}}(2-\mathcal{Y}_{eff}(l))^{\frac{1}{2}}log^{\frac{1}{2}}(\frac{r}{a_0})\label{ssz-semihel},
\end{align}
where $\mathcal{Y}_{eff}=2-\bar{\mathcal{Y}}_{int}$. At this superhelical-Zeeman boundary, the superconducting coupling tends toward strong coupling, and the dominant superconducting phase undergoes additional logarithmic corrections due to the Zeeman field. As a result, SDW correlations in the helical superconductor are suppressed by the Zeeman effect, while the Zeeman field incrementally enhances the pairing correlations, stabilizing a stronger helical topological superconducting phase.

\subsubsection{Zeeman-Induced Corrections to Short-Range Correlations in the Superhelical Edge}

The corrections to the short-range components of the correlation functions induced by the marginal Zeeman gap can be approximated in frequency space as
\begin{align}
&\mathcal{R}^{short-range}_{CDW}(\omega)\approx\omega +\mathcal{Y}_z^2\omega^{2(\bar{K}-2)},
\\
&\mathcal{R}^{xx, short-range}_{SDW}(\omega)\approx\omega+\mathcal{Y}_z^2\omega^{2\bar{K}-3}.
\end{align}
At low frequencies, the Zeeman term contributes significantly to the leading behavior on the charge phase only when $\bar{K} < 5/2$. In this situation, increasing the superconducting gap in the presence of repulsive interactions drives $\bar{K}$ to smaller values, which deepens the Zeeman gap. As a result, the enhancement of the superconducting gap makes the Zeeman-induced corrections to the charge sector more pronounced, making this phase more competitive. In the $x$-phase, a notable correction appears when $\bar{K} < 2$. This analysis demonstrates that, under short-range interactions, the charge density wave phase can strongly compete with the long-range correlations discussed earlier.

\section{Spin Transport in the Magnetized Superconducting Helical Edge} \label{s6}

In this section, we investigate the spin transport properties of the helical edge state in the presence of both Zeeman and superconducting gaps. The Zeeman gap modifies the carrier velocity and, due to the helical feature of the states, mixes the spin state, in particular near $k = 0$. In the presence of superconducting pairing, the additional spin gap opens through the sine-Gordon term which depends on the spin bosonic field $\Theta$ that may impact spin transport. In what follows, in order to analyze these effects, we calculate the spin conductivity employing the Memory function formalism \cite{Giamarchi1991}. Furthermore, to capture behavior over a broad temperature range, we present RG-improved Memory function results.

Before proceeding with interaction effects, it is useful to recall the noninteracting limit of the helical edge described by Eqs. (\ref{Ham1})–(\ref{Ham1Sup}). Diagonalizing the single-particle Hamiltonian
$H_{\mathrm{hel}} + H_{z} + H_{s}$ and evaluating the Kubo formula in the absence of electron–electron interactions yields a quantized ballistic spin conductance for a gapless helical edge. While a finite Zeeman or superconducting gap produces an activated suppression of spin transport: for $T \gg \Delta$ the spin conductance remains close to its quantized value, whereas for $T \lesssim \Delta$ it is strongly reduced due to the opening of a spin gap in the excitation spectrum (see Appendix \ref{app:NonIntGs}). In the following, the memory-function and RG-improved results should be viewed as interaction-induced power-law renormalizations of this simple crossover: repulsive interactions weaken the temperature dependence and stabilize nearly ballistic spin transport, while attractive interactions enhance the effective gap and lead to a stronger suppression of the spin conductivity with decreasing temperature.

\subsection{The frequency dependence of spin conductivity}

Based on the Kubo formula for conductivity~\cite{Mahan}, the conductivity can be written as~\cite{Giamarchi1991}
\begin{equation}
\sigma(\omega, T) = \frac{i \tilde{v}}{\pi \tilde{K}} \frac{1}{\omega + M(\omega, T)},
\label{Condct}
\end{equation}
where
\begin{equation}
M(\omega, T) = \frac{\omega \chi(\omega, T)}{\chi(0, T) - \chi(\omega, T)},
\label{Memo}
\end{equation}
is the Memory function, and $\chi(\omega, T)$ is the current-current correlation function. Following Ref. \cite{Giamarchi2004}, we find the Memory function at $T = 0$ as
\begin{equation}
\begin{split}
M(\omega, 0) \simeq \frac{1}{\tilde{K}} \Delta^2 \sin(\pi \tilde{K}^{-1}) \Gamma^2(1 - \tilde{K}^{-1}) \\
\times e^{-i \pi (\tilde{K}^{-1} - 1)} \frac{1}{\omega} \left( \frac{a_0 \omega}{2 \tilde{v}} \right)^{2 \tilde{K}^{-1} - 2}.
\end{split}
\label{wmemory}
\end{equation}
Substituting Eq. (\ref{wmemory}) into Eq. (\ref{Condct}) shows that the frequency dependence of the conductivity takes the form
\begin{equation}
\sigma(\omega, T = 0) = \frac{i \tilde{v}}{\pi \tilde{K}} \frac{1}{\omega + \mathrm{Re}\, M(\omega, 0) + i\, \mathrm{Im}\, M(\omega, 0)}.
\label{Condct1}
\end{equation}
By taking the real part of the conductivity, one arrives at
\begin{equation}
\mathrm{Re}\, \sigma(\omega) = \mathcal{D} \, \delta(\omega) + \sigma_{\mathrm{reg}}(\omega),
\end{equation}
where $\mathcal{D}$ is the stiffness parameter and the regular part is given by
\begin{align}
\sigma_{\mathrm{reg}}(\omega) \propto \Delta^{2} \, \omega^{2 \tilde{K}^{-1} - 5}.
\label{Tconduct}
\end{align}

\subsection{The temperature dependence of spin conductivity}  

Using a similar approach to the previous section and following Refs.~\cite{Giamarchi2004,Visuri2020}, we find the spin conductivity at $\omega = 0$ as
\begin{align}
\sigma(\omega=0, T) \propto \Delta^{-2} T^{3 - 2 \tilde{K}^{-1}}.
\label{Tconduct}
\end{align}
One can explicitly see that the temperature dependence of the conductivity follows a nonuniversal power law, with the exponent controlled by interactions and the Zeeman gap, in contrast to other works.

\begin{figure}[t!]
    \centering
    \includegraphics[width=8cm]{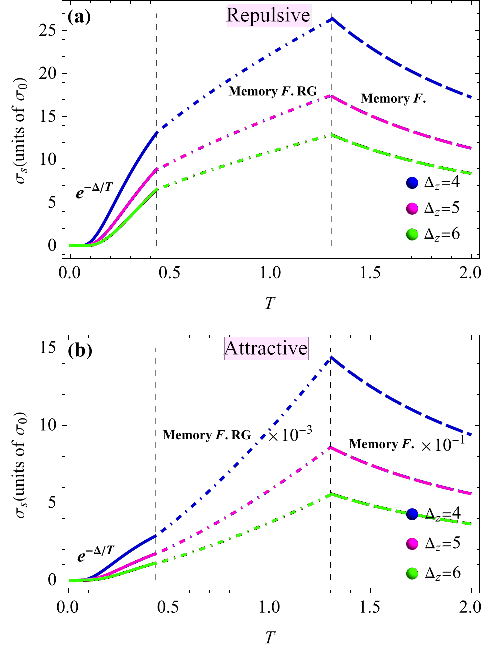}
\caption{
(Color online) Spin conductivity (in units of the conductance quantum) as a function of temperature for (a) repulsive and (b) attractive interactions, with $\Delta_s = 2$ and $\Delta_z = 4,\,5,\,6$. Three temperature regimes are distinguished: high temperatures ($T \gg \Delta$, dashed lines), temperatures on the order of the gap ($T \gtrsim \Delta$, dot-dashed lines), and low temperatures ($T < \Delta$, solid lines). The conductivity is evaluated via the standard memory function method, RG-improved memory function, and exponential decay formula, respectively. For clarity, values in each interval are slightly adjusted to ensure continuity at the regime boundaries.
}    
\label{fig8}
\end{figure}

It is important to note that the unusual power-law behavior obtained for the spin conductivity is valid only in the high-temperature limit. To extend our results to lower temperature range, we invoke the RG analysis. Considering the RG flow of the gap parameter in Eq. (\ref{YRG}), this parameter scales as
\begin{equation}
\mathcal{Y}(l) = \mathcal{Y}(0) \, e^{(2 - \tilde{K}^{-1}(l))l}.
\end{equation}
Here, we assume that $\tilde{K}(l)$ varies slowly with $l$. In this scenario, temperature serves as the energy cutoff, so the renormalization should proceed down to a cutoff of the order of the thermal length, $L_T \sim \frac{v}{T}$, corresponding to $e^{l^*} \sim \frac{L_T}{a_0}$. By stopping the scaling at $L_T$, we use the renormalized values $\Delta(L_T)$, $\tilde{K}(L_T)$, and $\tilde{v}(L_T)$ as the temperature-dependent quantities in the Memory function and in the conductivity, instead of the bare parameters $\Delta$, $\tilde{K}$, and $\tilde{v}$. The RG-improved expression for the conductivity then reads
\begin{equation}
\begin{split}
\sigma(L_T) =& \frac{\tilde{v}(L_T)}{16 \pi^4 T} \mathcal{Y}(L_T)^{-2} \mathbf{B}^{-2} \left( \frac{\tilde{K}^{-1}(L_T)}{2}, 1 - \tilde{K}^{-1}(L_T) \right) \\
&\times \cos^{-2} \left( \pi \frac{\tilde{K}^{-1}(L_T)}{2} \right),
\label{sigmaLT}
\end{split}
\end{equation}
where $\mathbf{B}(x, y)$ is the Beta function. Note that, at high temperatures, the gap coefficient acts as a constant value and does not undergo rescaling.

Numerically evaluated spin conductivity versus temperature is shown in Fig. \ref{fig8}(a) and \ref{fig8}(b) for repulsive and attractive interactions, respectively. Each diagram displays the spin conductivity of the super-PMH edge across three temperature regimes: the high-temperature regime ($T \gg \Delta$, dashed lines), the intermediate regime just above the gap ($T \gtrsim \Delta$, dot-dashed lines), and the low-temperature regime ($T < \Delta$, solid lines). Conductivities in these regimes are obtained using the standard memory function method, RG-improved memory function, and an exponential decay model for electrons, respectively. All quantities are shown in dimensionless form, with the Boltzmann constant $k_B$ set to unity.

In both cases, the spin conductivity increases with decreasing temperature when $T \gg \Delta$. As $T$ approaches $\Delta$, the gap coefficient $\Delta = \Delta_s \sin[2\vartheta_{k_F}]$ becomes significant, leading to a marked reduction in conductivity. For $T < \Delta$, the memory function approach is no longer valid and the conductivity of free quasiparticles is exponentially suppressed as $\sigma_s(T) \propto e^{-\Delta/T}$ \cite{Visuri2020,Hsu2018,Rice1976,Nattermann2003}.

Two further trends are apparent: first, the spin conductivity decreases systematically with increasing Zeeman field in both regimes. Second, for attractive interactions [Fig.~\ref{fig8}(b)], the Zeeman gap substantially enhances the superconducting gap, resulting in a much sharper drop in conductivity—approximately three orders of magnitude lower than in the repulsive case [Fig.~\ref{fig8}(a)] at temperatures of the order of the gap.

\subsection{Spin Conductance}

By considering the length parametric regime $L_T \ll L_{\Delta}, L$, we calculate the spin conductance of a finite edge connected to non-interacting leads in the vicinity of conventional superconductivity. Since the leads introduce resistance at the contact with the 1D edge, this value must be included in the calculation of the total conductance, in addition to the resistance of the wire itself \cite{Visuri2020}:
\begin{equation}
    R_T = R_{\text{edge}} + R_{\text{contact}}.
\end{equation}
The spin conductance is then given by
\begin{equation}
    G_{s} = \frac{1}{R_T} \approx 1 - R_{\text{edge}},
\end{equation}
where we have set $R_{\text{contact}} = 1$ and used the approximation that $R_{\text{edge}}$ is very small. The resistance is related to the conductivity by the specific resistivity $R = \rho L$, where $\rho = 1 / \sigma$. This yields the following expression for the conductance:
\begin{equation}
    G_{s} - 1 \propto -\mathcal{Y}^2\frac{L}{a_0} (\frac{2\pi a_0 T}{\tilde{v}})^{2 \tilde{K}^{-1} - 3}.
    \label{Sconductance}
\end{equation}

\begin{figure}[t!]
    \centering
    \includegraphics[width=8cm]{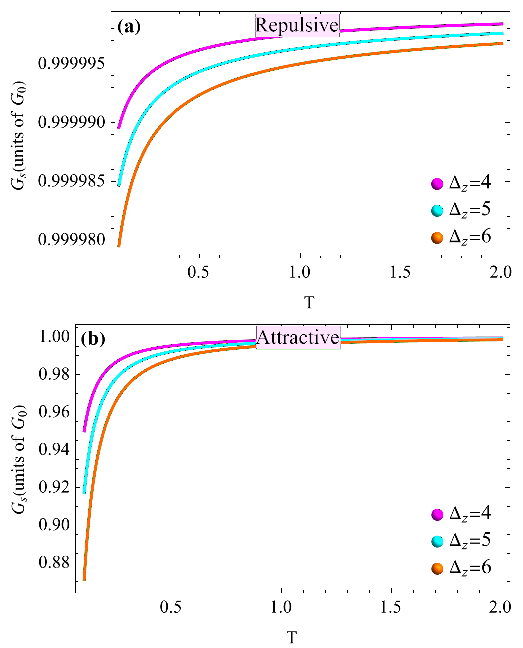}
    \caption{(Color online) Spin conductance (in units of the conductance quantum) as a function of temperature for (a) repulsive and (b) attractive interactions, with $\Delta_s = 2$ and $\Delta_z = 4,\,5,\,6$. All quantities are shown in dimensionless form, with the Boltzmann constant $k_B$ set to unity.}
\label{fig9}
\end{figure}

It can be seen that when $\tilde{K}^{-1} > 3/2$, decreasing the temperature reduces the correction of the conductance. In contrast, when $\tilde{K}^{-1} < 3/2$, the gap term becomes relevant in the RG sense, and lowering the temperature increases the correction, resulting in a decrease in the conductance of the super-PMH system. This shows that strong attractive interactions significantly enhance the effectiveness of the superconducting gap. In this regime, increasing the Zeeman gap reduces the threshold of interaction strength needed for the superconducting gap to affect the conductance correction. This means that since a stronger Zeeman field in the presence of attractive interactions further enhances the superconducting gap, even relatively weak attraction can produce the same impact on conductance as would otherwise require much stronger interactions. As discussed in Section \ref{V-A-2}, for $3/2 < \tilde{K}^{-1} < 2$, the system experiences weak repulsion, and the edge state is in a relevant gap regime. However, in this case, the superconducting gap does not reduce the conductance (Eq. (\ref{Sconductance})). Moreover, increasing the Zeeman gap further broadens the regime where the superconducting gap is ineffective in reducing the conductance, making successful conductance correction possible just at much weaker electron repulsion.

Figure \ref{fig9} illustrates the temperature dependence of the spin conductance for the super-PMH edge in various interaction regimes. In panels (a) and (b), we show the normalized spin conductance $G_s$ as a function of temperature for repulsive ($\widetilde{K}^{-1} > 1$) and attractive ($\widetilde{K}^{-1} < 1$) interactions, respectively, for various values of the Zeeman gap. For strong repulsive interactions $\widetilde{K}^{-1} > 3/2$ (see Fig. \ref{fig9}(a)) the conductance remains close to its quantized value as temperature decreases, reflecting the irrelevance of the superconducting gap and nearly ideal spin transport. In contrast, as the interaction becomes more attractive $\widetilde{K}^{-1} < 3/2$ (see Fig. \ref{fig9}(b)), the superconducting gap term becomes relevant in the renormalization group sense. In this regime, lowering the temperature leads to a pronounced suppression of spin conductance, consistent with the opening of a spin gap in the excitation spectrum.

Importantly, increasing the Zeeman gap in the presence of attractive interactions lowers the critical interaction strength necessary for the superconducting gap to dominate. As a result, even relatively weak attractive interactions can substantially suppress the spin conductance, similarly to the effect produced by stronger attraction at lower Zeeman fields. This interplay is manifested in the shifting conductance curves in Fig. \ref{fig9} as functions of both the interaction and Zeeman field.

\section{Summary and Conclusions}
\label{s7}

In this work, we have investigated the interplay between electronic correlations, Zeeman coupling, and s-wave superconducting proximity in 1D helical edge states. Employing bosonization and RG methods, we have analyzed how the competition between magnetic and superconducting gaps, together with electron-electron interactions, determines the dominant ordering tendencies and correlation functions in both the PMH and superhelical regimes.

Our analysis shows that repulsive interactions generally favor magnetic order, while attractive interactions enhance superconducting pairing. The relevance or irrelevance of the Zeeman and pairing gaps, imposed by the Luttinger parameter, leads to a rich phase diagram in which the dominant phase can be tuned by controlling the interaction strength, the Zeeman field, or the superconducting proximity.

We have derived analytical expressions for the density and spin correlation functions, identifying regimes where CDW, SDW, singlet, or triplet pairing correlations dominate. In particular, we have shown that logarithmic corrections, arising from the competition between marginal or nearly-marginal gap terms, can substantially modify the scaling of correlation functions and influence the stability of different phases. We further analyzed the corrections to short-range correlations using perturbation theory, and clarified the conditions under which the charge sector or triplet superconducting order can become dominant. Our results indicate that the coexistence of CDW and triplet pairing correlations can be stabilized by appropriately tuning the Zeeman gap and electron-electron interactions.

The spin transport properties of the helical edge is studied via the Kubo and Memory function approaches. We found that the frequency and temperature dependence of the spin conductivity is influenced by the interaction-dependent exponents. It is shown also that the RG flow of the gap parameters leads to significant renormalization of the low-energy transport. Our results for the spin conductance of a finite edge connected to non-interacting leads also indicate the role of contact resistance and show how superconductivity and magnetism along with mediating interactions can combine to modify the observable response.

Finally, we comment on the possibility of numerically benchmarking our results. The present analysis relies on bosonization and renormalization group methods, which are controlled in the low-energy limit. A natural complementary approach is provided by density matrix renormalization group (DMRG) simulations applied to lattice realizations of interacting helical edges with spin–orbit coupling, Zeeman field, and proximity-induced pairing. Such models can access the same parameter regimes considered here (weak-to-moderate interactions and competing magnetic and superconducting gaps) and allow direct evaluation of correlation functions and their scaling behavior, offering a quantitative test of our predictions for CDW, SDW, and pairing instabilities \cite{White1992,Schollwöck2005,Stoudenmire}. More broadly, tensor-network methods such as projected entangled pair states (PEPS) could incorporate the two-dimensional bulk and edge simultaneously in geometries such as cylinders or disks, enabling a study of how bulk topology influences edge correlations in the partially mixed helical regime \cite{Verstraete2004,Orus2014}. While these approaches are computationally demanding, they provide a promising route for future investigations of the interplay between interactions, magnetism, and superconductivity in topological systems.

\section*{Data Availability Statement}
The data that support the findings of this study are available from the corresponding author upon reasonable request.

\section*{Acknowledgement}
We are grateful to B. Braunecker for valuable discussions.
\appendix
\begin{widetext}
\section{Noninteracting spin conductance baseline}
\label{app:NonIntGs}

In this appendix we summarize the noninteracting spin-transport behavior of the magnetized, proximitized helical edge, which serves as a reference for the interacting results discussed in Sec. \ref{s6}.

In the presence of Zeeman field $\Delta_z$ and proximity-induced pairing $\Delta_s$, the single-particle dispersion reads
\begin{equation}
  \epsilon_{\pm,\eta}(k_x) = \eta\,\sqrt{v_F^2 k_x^2 + (\Delta_z\pm\Delta_s)^2}, 
  \qquad \eta = \pm 1.
\end{equation}
For a Fermi level at $\mu=0$, the minimal positive excitation energy occurs at $k_x = 0$, where the four energies are
\begin{equation}
  E_1 = \Delta_s + \Delta_z,\quad
  E_2 = \Delta_s - \Delta_z,\quad
  E_3 = -\Delta_s + \Delta_z,\quad
  E_4 = -\Delta_s - \Delta_z.
\end{equation}
The smallest positive energy is therefore
\begin{equation}
  \Delta_{\text{eff}} = \min\{E_i>0\} 
  = \bigl|\,|\Delta_z| - |\Delta_s|\,\bigr|,
\end{equation}
so that the combined action of Zeeman and pairing reduces to an effective single-particle gap $2 \bigl|\,|\Delta_z| - |\Delta_s|\,\bigr|$ at the Fermi level. The gap closes when $|\Delta_z| = |\Delta_s|$, signaling the transition between a Zeeman-dominated and a pairing-dominated regime.

For a noninteracting 1D helical edge, the linear-response spin conductance can be written in the form \cite{NazarovBlanter2009}
\begin{equation}
  G_s(T) = G_{0} \int_{-\infty}^{\infty} dE\,
  \bigl(-\partial_E f(E,T)\bigr)\, \mathcal{S}(E),
\end{equation}
where $G_{0}$ is the spin conductance quantum, 
$f(E,T) = 1/[\mathrm{e}^{E/k_B T}+1]$ is the Fermi function, and $\mathcal{S}(E)$ is the spin polarity.

In the presence of the hard gap $\Delta_{\text{eff}}$ at the Fermi level and otherwise, we approximate
\begin{equation}
  \mathcal{S}(E) =
  \begin{cases}
    0, & |E| < \Delta_{\text{eff}},\\[2pt]
    1, & |E| > \Delta_{\text{eff}}.
  \end{cases}
\end{equation}
Then
\begin{equation}
  \frac{G_s(T)}{G_0} =
  \int_{|E|>\Delta_{\text{eff}}} dE\,\bigl(-\partial_E f(E,T)\bigr).
\end{equation}
Using $\int_{-\infty}^{\infty} dE\,(-\partial_E f) = 1$ and splitting the integral, we obtain
\begin{equation}
  \frac{G_s(T)}{G_0} 
  = 1 - \int_{-\Delta_{\text{eff}}}^{\Delta_{\text{eff}}} dE\,
  \bigl(-\partial_E f(E,T)\bigr)
  = 1 - \bigl[f(-\Delta_{\text{eff}}) - f(\Delta_{\text{eff}})\bigr].
\end{equation}
Since $f(-E) = 1 - f(E)$, this simplifies to
\begin{equation}
  \frac{G_s(T)}{G_0} 
  = 2 f\!\left(\frac{\Delta_{\text{eff}}}{k_B T}\right),
\end{equation}
with $f(x) = 1/(\mathrm{e}^x+1)$.

In particular, for the gapless helical edge ($\Delta_{\text{eff}}=0$) with $f(0)=1/2$, one gets
  \begin{equation}
    \frac{G_s(T)}{G_0} = 1,
  \end{equation}
  i.e., a quantized, temperature-independent noninteracting spin conductance. On the other hand, for the gapped edge ($\Delta_{\text{eff}}>0$) and $k_B T \ll \Delta_{\text{eff}}$, one obtains
  \begin{equation}
    f\!\left(\frac{\Delta_{\text{eff}}}{k_B T}\right)
    \simeq \mathrm{e}^{-\Delta_{\text{eff}}/k_B T},
  \end{equation}
  so that
  \begin{equation}
    \frac{G_s(T)}{G_0} \simeq 2\,\exp\!\left(-\frac{\Delta_{\text{eff}}}{k_B T}\right),
  \end{equation}
  i.e., an activated suppression of spin transport at low temperature.
  
The resulting temperature dependence of the noninteracting spin conductance is illustrated in Fig. \ref{fig:nonintGs}. For a gapless edge $(\Delta_z=\Delta_s=0)$ the spin conductance remains quantized and independent of temperature. As soon as a finite effective gap  $\Delta_{\mathrm{eff}} = ||\Delta_z|-|\Delta_s||$ opens, $G_s(T)$ is strongly suppressed at low temperatures and crosses over towards the quantized value at  $T \gtrsim \Delta_{\mathrm{eff}}$. Close to the line $\Delta_z \approx \Delta_s$ the effective gap is small and the conductance stays nearly quantized over a broad temperature range. While increasing $||\Delta_z|-|\Delta_s||$ enhances the effective gap and shifts the crossover $T \sim \Delta_{\text{eff}}/k_B$ to higher temperature leading to a stronger suppression of the conductance. This provides the reference behavior against which the interaction-induced power laws and RG-improved results of Sec. \ref{s6} are to be compared.

\begin{figure}[t]
  \centering
  \includegraphics[width=0.6\columnwidth]{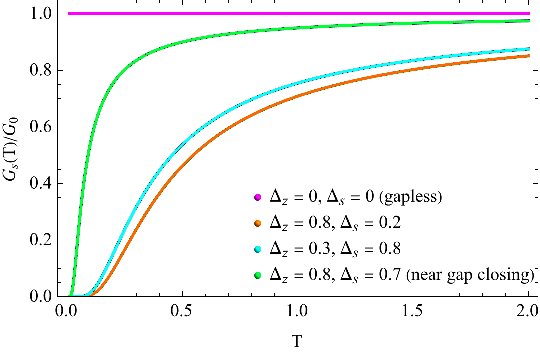}
  \caption{(Color online) Noninteracting spin conductance 
    $G_s(T)/G_0$ as a function of temperature $T$ for different
    values of Zeeman and pairing gaps $(\Delta_z,\Delta_s)$.}
  \label{fig:nonintGs}
\end{figure}
\end{widetext}


\end{document}